\documentclass[a4paper,fleqn,usenatbib]{mnras}

\usepackage{newtxtext,newtxmath}
\usepackage{mathptmx}
\usepackage[T1]{fontenc}
\usepackage{ae,aecompl}
\usepackage{graphicx}	% Including figure files
\graphicspath{{fig/}}
\usepackage{amsmath}	% Advanced maths commands
\usepackage{amssymb}	% Extra maths symbols
\usepackage{multicol}        % Multi-column entries in tables
\usepackage{bm}		% Bold maths symbols, including upright Greek
\usepackage[flushleft]{threeparttable} % table notes

\title[BeSOS survey]{Stellar parameters and H$\alpha$ line profile variability of Be stars in the BeSOS survey}

\author[C. Arcos et al.]{
C. Arcos,$^{1}$\thanks{E-mail: catalina.arcos@uv.cl}
S. Kanaan,$^{1}$
J. Chavez,$^{1}$
L. Vanzi,$^{2}$
I. Araya$^{1}$ 
and M. Cur\'e$^{1}$
\\
$^{1}$Instituto de F\'isica y Astronom\'ia, Facultad de Ciencias, Universidad de Valpara\'iso.  Av. Gran Bretana 1111, Valpara\'iso, Chile.\\
$^{2}$Department of Electrical Engineering and Center of Astro Engineering, Pontificia Universidad Cat\'olica de Chile, \\
Av. Vicuna Mackenna 4860, Santiago, Chile.\\
}

\date{Accepted XXX. Received YYY; in original form ZZZ}

\pubyear{2017}

\begin{document}
\label{firstpage}
\pagerange{\pageref{firstpage}--\pageref{lastpage}}
\maketitle

% Abstract of the paper (<250 words)
\begin{abstract}
The Be phenomenon is present in about 20$\%$ of the B-type stars. Be stars show variability on a broad range of timescales, which in most cases is related to the presence of a circumstellar disk of variable size and structure. For this reason a time resolved survey is highly desirable in order to understand the mechanisms of disk formation which are still poorly understood. In addition, a complete observational sample would improve the statistical significance of the study of the stellar and disk parameters.
The ``Be Stars Observation Survey'' (BeSOS) is a survey containing reduced spectra obtained using the echelle spectrograph PUCHEROS with a spectral resolution of 17000 in the range of 4260-7300 $\text{\AA}$. BeSOS's main objective is to offer consistent spectroscopic and time resolved data obtained with one instrument. The user can download or plot the data and get the stellar parameters directly from the website.  We also provide a star-by-star analysis based on photometric, spectroscopic and interferometric data as well as general information about the whole BeSOS sample. Recently, BeSOS led to the discovery of a new Be star \hbox{HD 42167} and facilitated study of the V/R variation of \hbox{HD 35165} and \hbox{HD 120324}, the steady disk of \hbox{HD 110335} and the Be shell status of \hbox{HD 127972}. Optical spectra  used in this work, as well as the derived stellar parameters are available online in \url{http://besos.ifa.uv.cl}.
\end{abstract}


%%%%%%%%%%%%%%%%%%%%%%%%%%%%%%%%%%%%%%%%%%%%%%%%%%

%%%%%%%%%%%%%%%%% BODY OF PAPER %%%%%%%%%%%%%%%%%%

\section{Introduction} \label{sec:intro}
Be stars are defined as ``non-supergiant Late O-type to early A-type star whose spectra have, or had at some time, one or more Balmer lines in emission". It is generally accepted now, that Be stars are very rapidly rotating main sequence stars, surrounded by a circumstellar envelope (CE) of gas. The angular momentum is transferred by some mechanism from the interior to the equatorial surface of the rotating star, where the disk is fed at a certain rate. This material is governed by viscosity and is placed in a thin equatorial disk with Keplerian rotation \citep{Meilland2007}. The model describing the physics of the disk is named ``Viscous Decretion Disk'' (VDD) and was proposed originally by \cite{Lee1991}. Moreover, Be stars show near-IR (1.0 - 2.5 $\mu$m) excess in the spectral energy distribution due to the free-bound and free-free emission from the same ionized circumstellar material that gives rise to the Hydrogen Balmer emission lines \citep[][]{Waters1987}. All these features refer to ``Classical Be stars''.

Observations of Be stars provide clear evidence that these stars are variable with timescales that range from minutes to years \citep{Porter2003}. Phase transitions are revealed by Balmer line emission intensity and global line profile changes as well as changes in the visible spectral energy distribution. Variations in the emission lines can occur over periods of days, months, or years. These changes can be due either to variations in the physical structure and size of a more or less permanent CE, or the creation of a new CE during mass ejection events of the central star. The origin of the Be phenomenon is still poorly understood. See \cite{Rivinius2013} for a complete review of Be stars.

In order to study the variability of Be stars on all timescales, a large amount of observation time is required in addition to a well defined observation strategy to constrain the variation that can go from hours to years. The variation is revealed in the changes that occur in the emission lines such as H$\alpha$ and H$\beta$ line profiles; not only the global line profile but also the violet (V) and red (R) peaks or the V/R ratio and the double peak separation (DPS). These variations are directly linked to the changes in the CE and analyzing them will help understand the origin of this phenomena. Observations in the southern hemisphere of Be stars brighter than visual magnitude of 8, are limited in comparison with northern Be stars in terms of number of spectra per object. For these reasons the creation of the ``Be Stars Observation Survey'' (BeSOS) is important (see the Appendix~\ref{Ap:besoshandbook} available as an online material, for details on the use of the website).

BeSOS not only tracks variations of Be stars but also aims to confirm and discover new Be stars by compiling a list of potential candidates using the photometric data in the literature in a search for IR excess. Moreover BeSOS is an homogeneous survey since the data are obtained using solely one instrument. In fact, in comparison with other databases, BeSOS uses one instrument, one spectral resolution and the same steps of data reduction, making it perfect for relative studies of Be stars mainly by comparing spectra in different epochs directly and detecting features in the spectra due to large or small variations within the spectral resolution.

\section{BeSOS: the data}

\subsection{The main instrument PUCHEROS}
The observations were carried out using the spectrograph PUCHEROS\footnote{\url{http://www2.astro.puc.cl/ObsUC/index.php/PUCHEROS}} developed at the Center of Astro-Engineering of Pontifica Universidad Catolica (PUC) \citep{Infante2010}. PUCHEROS is installed at the ESO 50 cm telescope of the Observatory UC (OUC) Santa Martina located near Santiago, Chile. The instrument is based on a classic echelle design connected to the telescope via a 50 micron core optical fiber. PUCHEROS covers the entire visible spectral range (3900-7300 $\text{\AA}$) in one single exposure, but due to the low efficiency in the blue, we cut this part of the spectra and the obtained range is from 4260 to 7300 $\text{\AA}$. The spectral resolution is near to $\sim$17000 and the conservative estimate of the limiting magnitude is V=9, with an exposure time of 1 hour, reaching a S/N = 20, in case of poor seeing. The spectra are recorded in Flexible Image Transport System (FITS) format. More details about the instrument are provided in \cite{Vanzi2012}.

\subsection{Target election and observation}
BeSOS observations started in November 2012. Our program stars come from a large list of confirmed southern Be stars taken from the BeSS database \citep{Neiner2011} and also from a list of B stars showing IR excess taken from the Simbad astronomical database \citep[][]{Wenger2000}. We then selected targets with a visual magnitude lower than 8 in order to achieve a high S/N ($\sim$ 100-200). Observations are planned every month if the weather conditions are favorable. 

The selection also follows the variability. More observation time is dedicated to targets known for showing variability and also when variability is detected, allowing BeSOS to give the best time coverage possible and necessary to constrain the variability.

%Our observations use 41 orders towards the red, spanning from 4260 to 7300 $\text{\AA}$ as mentioned before. 
A typical observation consists of dark, bias and flat frames taken at the end of the observation, science stars during the night and Th-Ar exposures every hour. The exposure time is chosen to reach a S/N in the range 150-200. For targets with exposure time over 10 minutes, the observation is divided into several exposures of 10 min and combined later into one frame.

\begin{figure}
   \centering
   \includegraphics[width=1\columnwidth]{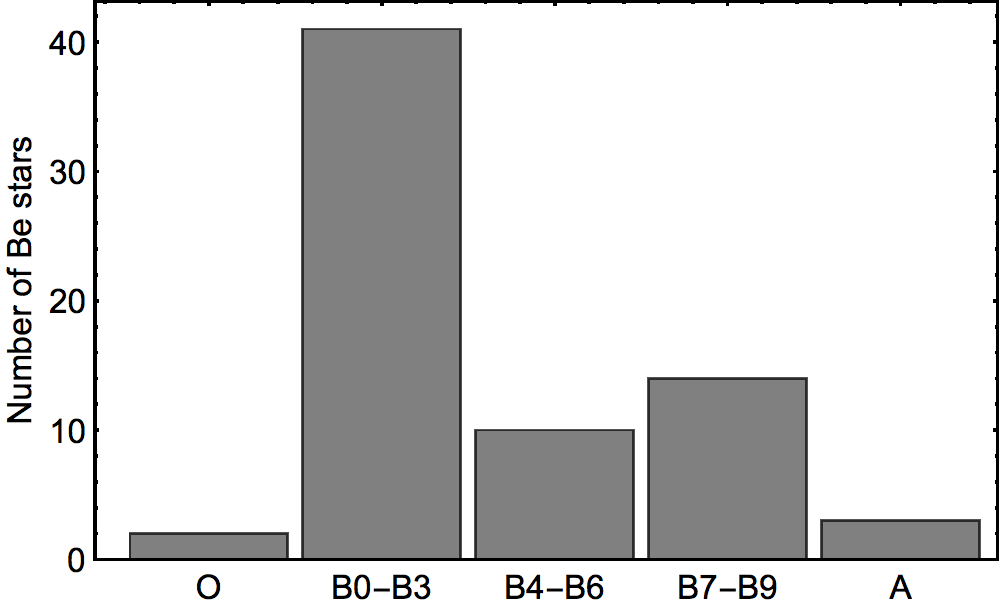}
   \caption{Number of Be stars vs spectral type from the literature in the BeSOS survey. The distribution peaks at early spectral types.}
   \label{Histo-Be}
\end{figure}

In Fig.~\ref{Histo-Be} we show the distribution of spectral type classification in the literature for Be stars in the BeSOS survey. For B-types, we separated the sample in early (B0-B3), mid (B4-B6) and late (B7-B9) subspectral types. The distribution of spectral types peaks at early types. The most frequent type found in the sample is B2 with $\sim$ 30$\%$, the same was found from other authors in previous works for B2-types (see \citealt{Zorec1997}).

\subsection{Data reduction}
We used the CERES pipeline developed by the Center of Astro-Engineering of PUC, for the reduction, extraction, and analysis of BeSOS spectra. Details of the implemented routines and reduction steps can be found in \cite{Brahm2017}. We compared the pipeline result with the standard Echelle reduction steps using IRAF \citep[][]{Tody1993}. An overview of these IRAF steps are enumerated as follows: 

\begin{enumerate}
	\item	Preparation of the raw spectra by removing the bias and dark contributions using the default tasks found in the package \textit{noao}, \textit{imred} and \textit{ccdred}. 
 
	\item   Extraction of the spectrum using the IRAF echelle data reduction package called \textit{echelle}. Using the task 					\textit{apall}, we can define the apertures and fit the background to extract the spectrum.
 
	\item 	Normalization is done by dividing the spectrum of each order by a fit of its continuum.
	
	\item   The wavelength calibration is made by fitting to each science spectra an already defined list of the Th-Ar lamp 					spectra and using the task \textit{refspec} where we choose the lamp spectrum taken before and after each science 					observation. 
  	
  	\item 	Finally, the multi-order spectra are merged by the task \textit{scombine}. 
   
    \item 	The heliocentric velocity correction is done using the task \textit{rvcorr}.

\end{enumerate}
The manual reduction completely validated the pipeline results that were assumed for the rest of the work. The pipeline's output contains 41 orders in wavelength and intensity, with no heliocentric velocity correction and without normalization with the continuum. We applied the IRAF steps (iii), (v) and (vi), to normalize, merge and correct the heliocentric velocity, respectively. These final spectra are uploaded to the BeSOS website. Each FITS file header contains: target name, observation date, Universal time, exposure time, AR, DEC and the heliocentric correction. We also offer the possibility for the user to reduce the data by himself by requesting the raw data by email.  

\section{Analysis by star}
\label{analysisstar}
The stellar parameters were obtained by fitting models to photometric and spectroscopic data. For each star, we did an analysis which is separated in three sections: Photometry, Spectroscopy and Interferometry. It is important to note that the Interferometry section is not available yet, because the data is under analysis and will be added in the future. Here we describe the method used in the Photometry and Spectroscopy sections, the models used, the confidence interval calculations and the algorithm implemented. 

\subsection{Photometry}
In order to derive stellar parameters, for each star we query Vizier\footnote{\url{http://vizier.u-strasbg.fr/vizier/sed/}} \citep{Vizier2000} for photometric data. VizieR provides access to the most complete library of published astronomical catalogs (16197 catalogs) and data tables available online organized in a self-documented database. The data usually covers the visual and IR spectral range and are fitted by a solar composition synthetic model atmosphere. Two databases of model atmospheres were used in this work: Kurucz\footnote{\url{ftp://ftp.stsci.edu/cdbs/grid/k93models/}} \citep{kurucz1979} and Tlusty\footnote{\url{http://nova.astro.umd.edu/}} \citep{Hubeny1995}. Kurucz models are Local Thermodynamic Equilibrium (LTE), plane-parallel, hydrostatic model atmospheres, for a wide range of metallicities, effective temperatures and gravities. There are 7600 stellar model atmospheres available in the Kurucz grid model. On the other hand, Tlusty is a grid containing 1540 metal line-blanketed, Non-LTE, plane-parallel, hydrostatic model atmospheres for the basic parameters of O and B type stars. We used as a first step Kurucz models to fit the Spectral Energy Distribution (SED), since the effective temperature of late B-type stars (e.g., B7-B9) is usually lower than $<$ 15000 K, which is out of range of the Tlusty grid. Solar metallicity and turbulence velocity, $v_\text{turb} =$ 2 km s$^{-1}$, were used for all fitted models. Table~\ref{tab:ATM} shows the range of parameters offered by the Kurucz and Tlusty grids. In our model fitting, we leave the effective temperature ($T_\text{eff}$), the surface gravity ($\log$ g) and the stellar radius ($R_{\star}$) as free parameters. The distance is taken from the literature as well as the interstellar reddening $E(B-V)=A_{v}/R_{v}$, where the measured value for the Milky Way is $R_{v} \sim$ 3.1 \citep{Schultz1975}. 

\begin{table}
    \caption{Grid of models used in this work for Kurucz and Tlusty model atmospheres.}
    \centering
      \begin{tabular}{c|c|c}
      \hline
      Model                         &  Kurucz/step       & Tlusty/step     \\
      \hline
      $T_\text{eff}$ (K)            & 3500-50000/250 & 15000-30000/1000\\    
      $\log$ g (cm s$^{-2}$)        & 0.0-5.0/0.5   & 1.75-4.75/0.25\\
      Solar Metallicity             & [M/H] = 0     & Z/Z$_{0}$ = 1.0\\
      $\lambda$ ($\mu$m)            & 0.0091 to 160 & 0.09 to 1 \\
      $v_\text{turb}$  (km s$^{-1}$)     & 2.0           &  2.0      \\
      \hline
      \end{tabular}
      \label{tab:ATM}
\end{table}

To find the best model we used the library IDL routine \texttt{mpfit} that searches for the best fit using the ``Levenberg-Marquardt'' (LM) algorithm.  The LM is  a  standard  technique  for solving  nonlinear least  squares  problems. Nonlinear least squares methods iteratively reduce the sum of the squares of the errors between the function and  the  measured  data  points  through  a  sequence  of  updates  to  parameter  values. The LM curve-fitting method is a combination of two minimization methods:  the gradient descent method and the Gauss-Newton method.  In the gradient descent  method,  the  sum  of  the  squared  errors  is  reduced  by  updating  the  parameters in the steepest-descent direction.  In the Gauss-Newton method,  the sum of the squared errors is reduced by assuming the least squares function is locally quadratic, and  finding  the  minimum  of  the quadratic. The LM  method  acts more like a gradient-descent method when the parameters are far from their optimal value, and acts more like the Gauss-Newton method when the parameters are close to their optimal value\footnote{Description taken from \url{http://people.duke.edu/~hpgavin/ce281/lm.pdf}}.  
If ``x'' represents values of the independent variable, ``y'' represents a measurement for each value of ``x'', and ``err'' represents the error in the measurements, then the deviates, $\Delta$, can be calculated as follows: 
\begin{equation}
\Delta = \frac{y-F(x)}{\text{err}},
\end{equation}
where F is the function representing the model. If the values of ``err'' are the 1-$\sigma$ uncertainties in ``y'', then $\chi^{2} = \sum \Delta^{2}$.  \texttt{mpfit} will minimize the $\chi^{2}$ value. Once the free parameters are found ($T_\text{eff}$, $\log$ g and $R_{\star}$), we take the values of $T_\text{eff}$ and $\log$ g as input values to start the same minimization method explained above, but now in the spectroscopic section (see sec.~\ref{spec}). Here, the parameters are constrained by fitting the HeI $\lambda$4471 and MgII $\lambda$4481 lines. The best fit model parameters are used again in the photometry section and fixed at the same values obtained with the spectroscopic section, leaving only $R_{\star}$ free. In this iterative way, we find the stellar parameters for all BeSOS spectra. 

\begin{figure}
    \centering
    \includegraphics[width=1\columnwidth]{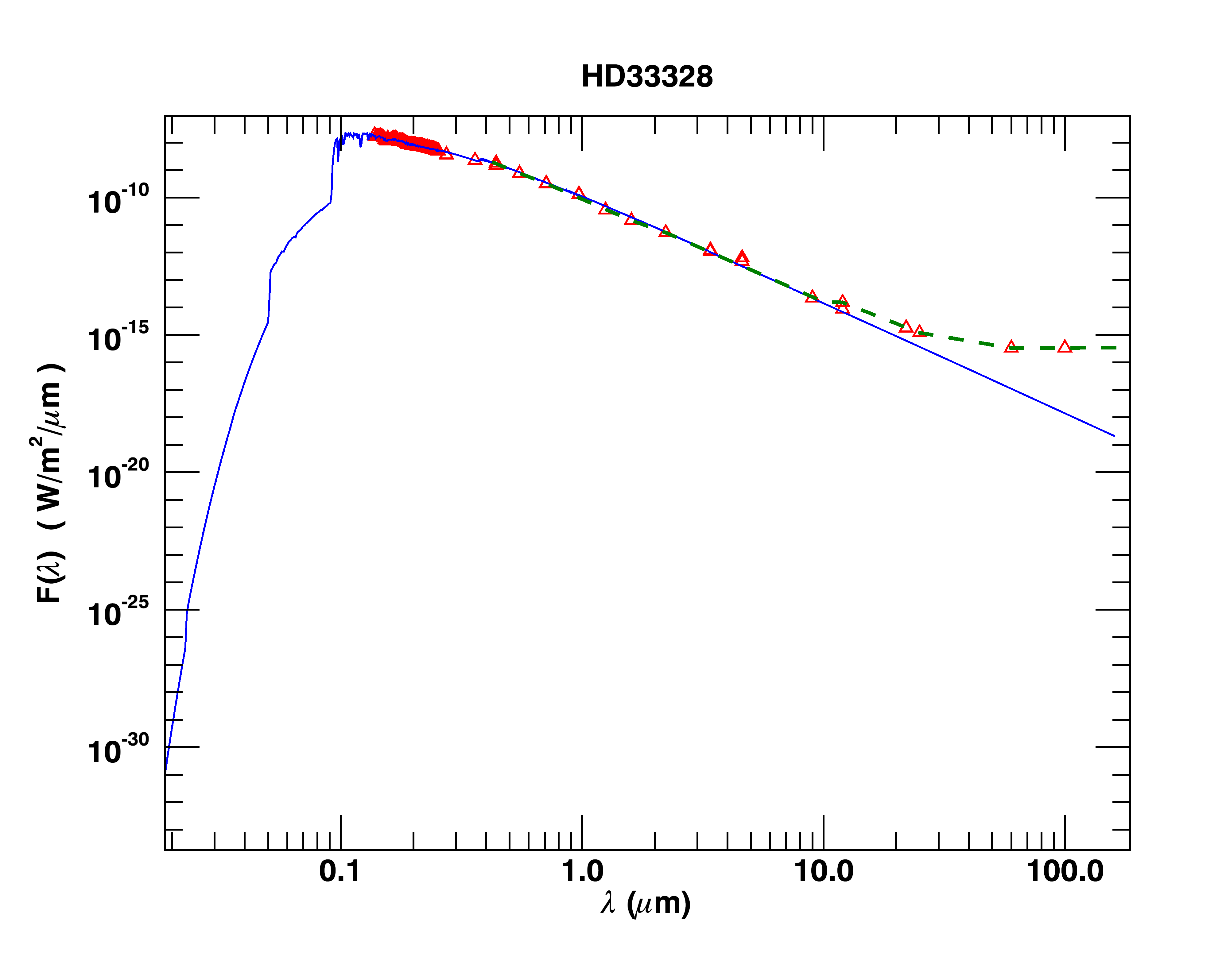}
    \caption{Example of the spectral energy distribution fit for the Be star HD 33328 using a Kurucz model. Empty red triangles are the photometric data, the solid blue line represents the best fitting Kurucz model and the dashed green line shows the infrared excess. The best fit is for a star with $T_\text{eff}$ = 19526 $\pm$ 195 K and  $\log$ g = 3.30 $\pm$ 0.03 cm s$^{-2}$.}
    \label{SED}
\end{figure}

An example of a spectral energy distribution model fitting is shown in Figure~\ref{SED} for the Be star HD 33328, where the user can see the data (red triangles), the best Kurucz model (solid blue line) and the infrared excess (dashed green line). The best fit parameters are: $T_\text{eff}$ = 19526 $\pm$ 195 K, $\log$ g = 3.30 $\pm$ 0.03 cm s$^{-2}$, $d$ = 248 pc, $R_{\star}$ = 7.31 $\pm$ 0.15 $R_{\odot}$ and E(B-V) = 0.04.  It is important to note that our actual photometric analysis does not fit the infrared excess. The reader is invited to use radiative transfer codes such as \textsc{Hdust} \citep{Carciofi2001} or \textsc{BEDISK} \citep{Sigut2007} to reproduce the IR excess as well as physical models of Be stars \citep[e.g.,][]{Domiciano2014,Silaj2016,Arcos2017}.

\subsection{Spectroscopy}
\label{spec}
\subsubsection{Calculations of stellar parameters}
Information about the stellar and disk parameters can be obtained by analyzing spectroscopic data. The gaseous disk is ionized due to the high temperatures of the central star. Therefore,  Balmer emission lines are seen at different intensities and with different shapes depending on the inclination of the disk and the spectral type of the central star \citep{Hanuschik1996a}.  %\textbf{\cite{Silaj2014} modeled the H$\alpha$ emission line of eight Be shell\footnote{``shell'' is used to indicate that the central depression is below the continuum} stars with inclination angles around  45$^{\circ}$, 65$^{\circ}$ and 70$^{\circ}$ and higher and they found that shell status is not always due to the edge-on sight of view.} 
Using radiative transfer codes, one can reproduce these emission lines to estimate the parameters of the star+disk system. On the other hand, information about the central star can be obtained by fitting a model atmosphere to Helium and metallic lines (e.g., HeI $\lambda$4026, 4471, SiII $\lambda$4128, 4130, MgII $\lambda$4481, among others), which in principle are not affected by the emission of the disk \citep{Collins1974}. Usually, Helium lines are used as indicators of the $T_\text{eff}$, $\log$ g and the projected stellar velocity $v \, \sin i$. 

\begin{figure}
\centering
\includegraphics[width=\columnwidth]{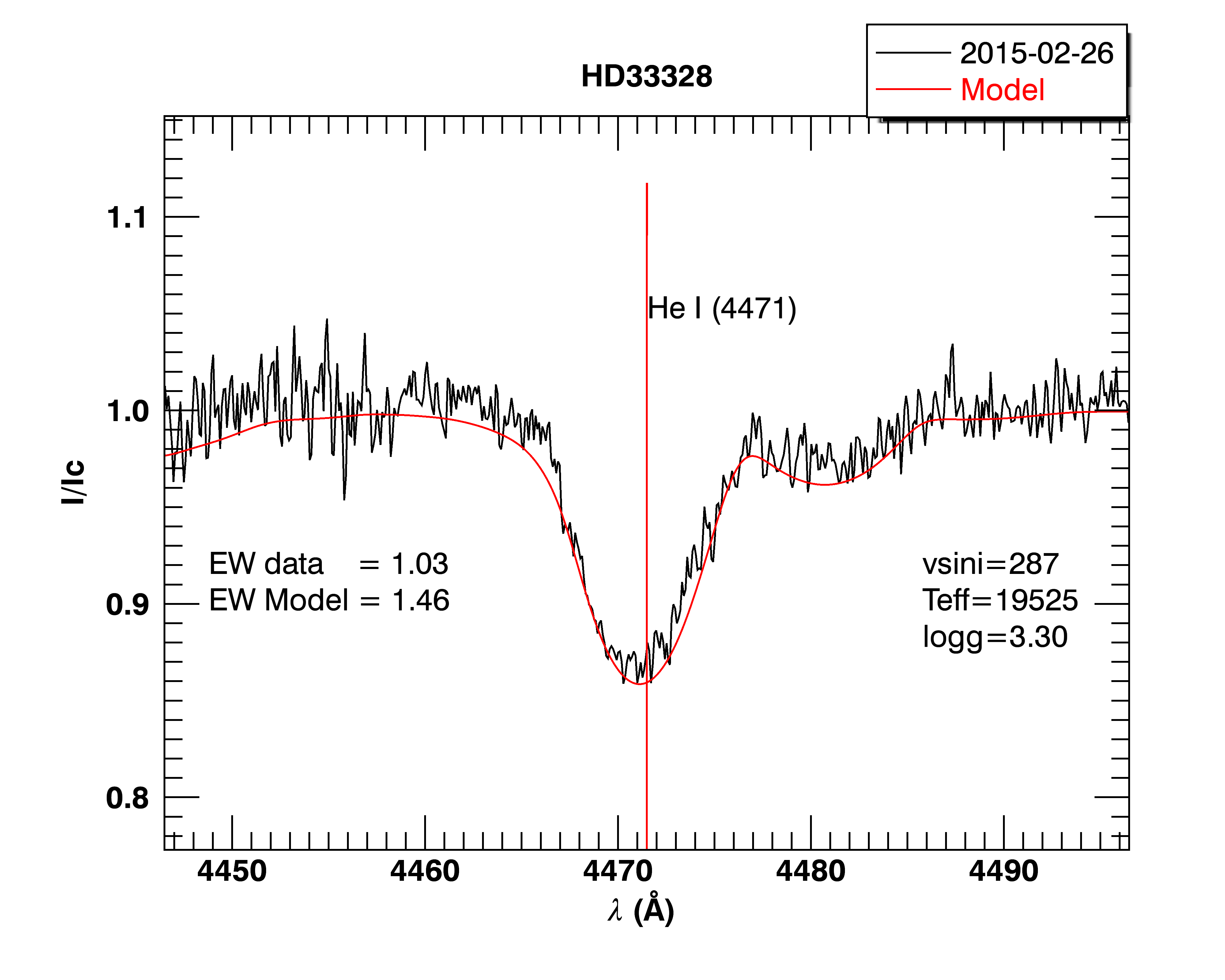}
 \caption{Example of the fit in the case of the Be star HD 33328 spectrum (solid black line) using a Tlusty model (solid red line) where the best parameters are displayed as a legend in the plot.}
   \label{tlusty-fit}
\end{figure}

As we explained in the photometry section, to constrain $T_\text{eff}$ and $\log$ g, we performed a fit to a rotationally convolved model on the HeI $\lambda$4471 and MgII $\lambda$4481 lines. These lines are always present in the Be star spectra, although depending on the sub-spectral type the intensities of the lines (depth) will be different. The variation of the ratio between HeI/MgII equivalent widths is used as an indicator of non-radial pulsations \citep{Vogt1990}. The best model was found using the \texttt{mpfit} IDL routine. The starting $T_\text{eff}$ and  $\log$ g values are from the least-square minimization of the atmosphere model and photometric data, and for $v \, \sin i$ the starting value is taken from the literature. The fit is made for the spectrum with the deepest HeI absorption line available in BeSOS for the selected target. We used Tlusty models in the spectroscopic section, because they are computed considering non-LTE and have more lines to resolve the statistical equations giving a better resolution than Kurucz models. Once we find the best fit parameters, we use them in the photometric and spectroscopic model fitting. An example of the best fit for the Be star HD 33328, is shown in the Figure~\ref{tlusty-fit}, where the data are shown in solid black line and the best model in  solid red line. The best parameters are shown in the legend of the plot, as well as the equivalent width of the data and the model.  

\subsubsection{Variability of the H$\alpha$ emission-line: disk information}
\begin{figure*}
\includegraphics[width=0.35\textheight]{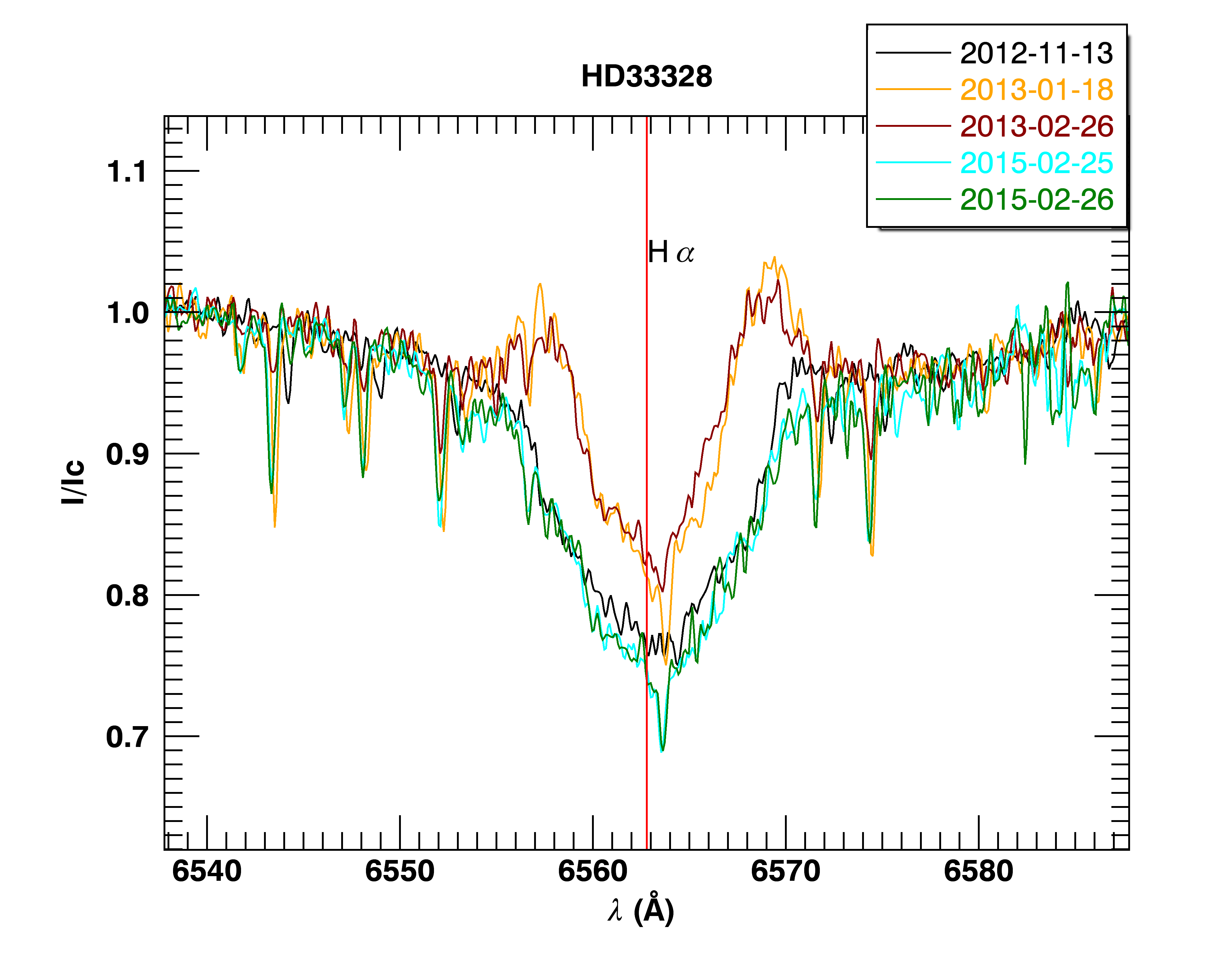}
\includegraphics[width=0.35\textheight]{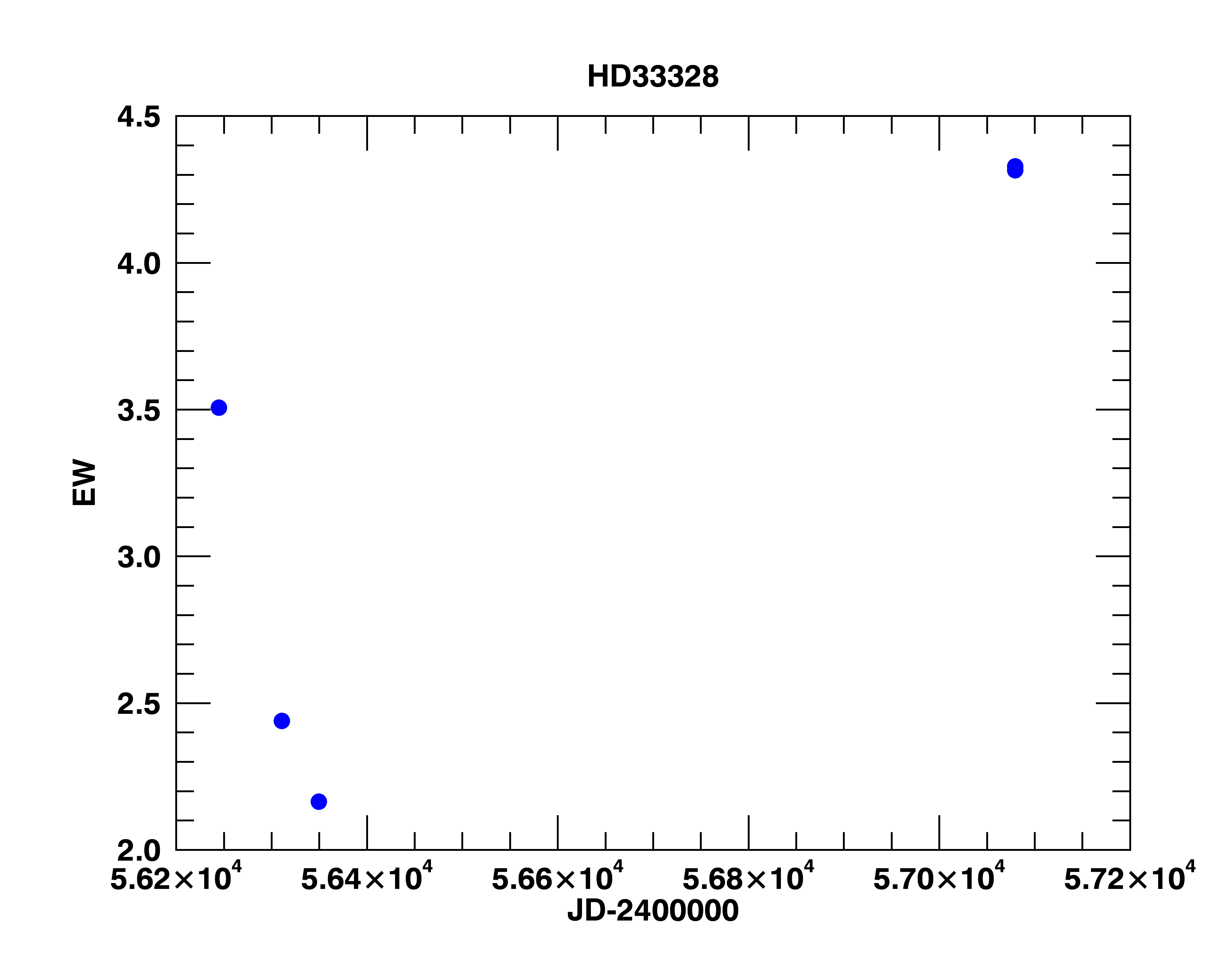}\\
\includegraphics[width=0.35\textheight]{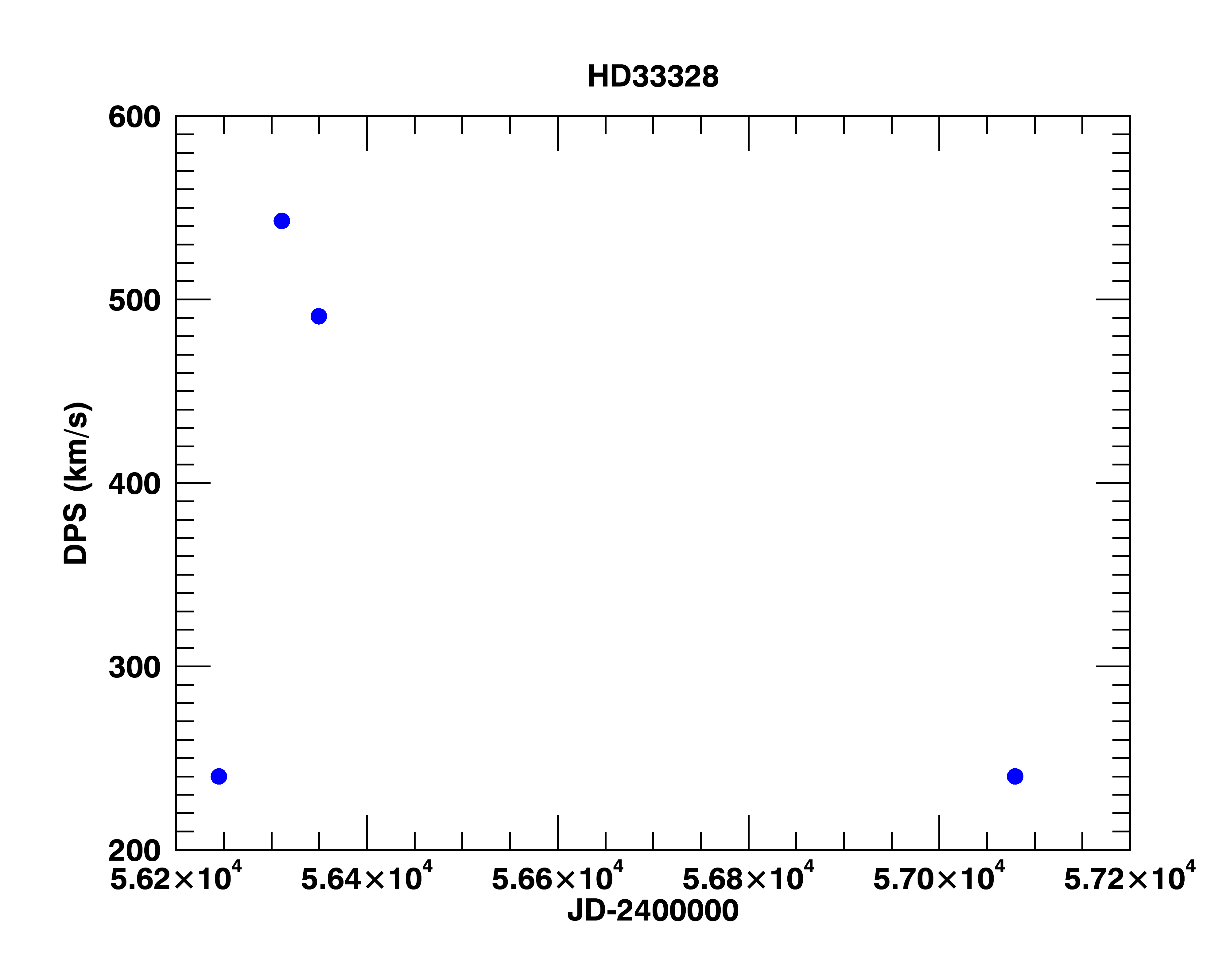}
\includegraphics[width=0.35\textheight]{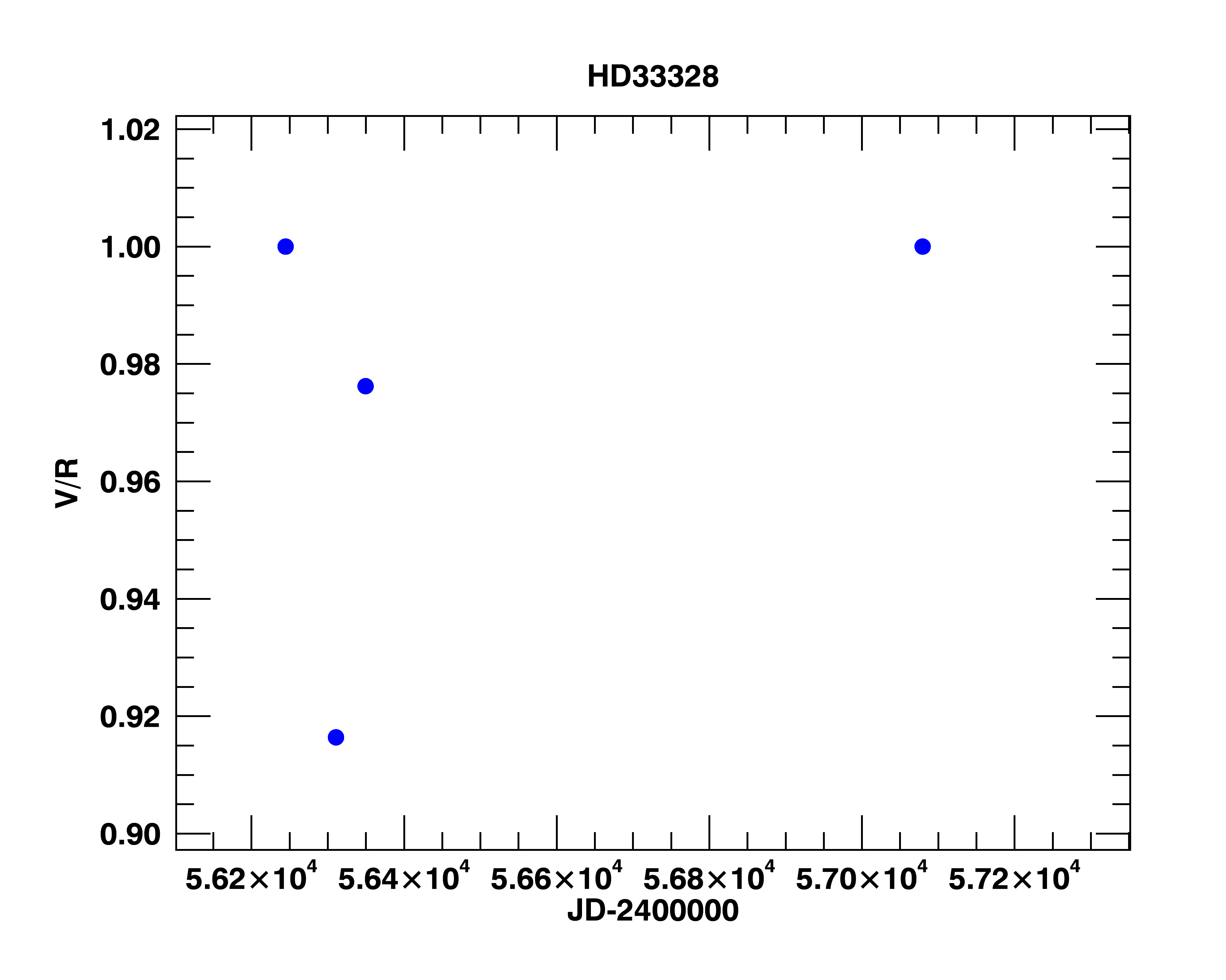}\\
\caption{Measurements of the H$\alpha$ emission line of the Be star HD 33328. 
The left upper plot shows all the spectra available for the star in the H$\alpha$ range. The upper-right plot shows the EW variation in Angstroms. The left bottom plot shows the DPS variation in km s$^{-1}$ and the right bottom plot shows the V/R variation.} 
\label{fig:Halpha-variation} 
\end{figure*}

As the intensity and the shape of the H$\alpha$ emission line is directly linked to the gaseous disk, analysis of the Double Peak Separation (DPS), the intensity of the Violet (V) and Red (R) peak, the V/R variation, the Radial Velocity (RV) and the Equivalent Width (EW) give information to estimate the rotation, symmetry, over-densities and size of the rotating disk, as well as information about binarity features if they exist. PUCHEROS's range (4260 - 7300 $\text{\AA}$) allow us to measure these mentioned features on three Balmer lines, H$\alpha$, H$\beta$ and H$\gamma$. Here, we present only measurements on the H$\alpha$ emission line, and in a future work we will add more analysis for the other emission lines. Therefore, for each star the spectroscopic section begins by showing a graph of H$\alpha$ for all spectra available in BeSOS, allowing the user to immediately check if the star exhibits variation. A second graph (if there is more than one spectrum) shows the H$\alpha$ EW (in Angstrom) over time (in Julian days), where the general convention was used for the sign, i.e, negative values are for emission lines (see Fig.~\ref{fig:Halpha-variation}). We also add the respective graph of the DPS (km s$^{-1}$), V/R and radial velocity (km s$^{-1}$).

The DPS and V/R are obtained by an interactive IDL program developed for this purpose (MUFIN : MUlti-FitINg). For each spectrum of a target the code searches for the V and R peaks, then the user must decide by visual inspection if the selected point corresponds to the V and R peak, respectively. If not, the user can move freely the interactive bars to fix the point. If the spectrum doesn't show a double peak line profile, then the user can click on either, ``1 peak profile'' or ``absorption profile''. For the option ``1 peak profile'' the DPS value is zero and for ``absorption profile'' the DPS value is set to the width of the H$\alpha$ wings of that spectrum. When going through all spectra available for the target, an ASCII file is generated for the DPS and V/R values versus Julian days, as well as their corresponding plots (see Fig.~\ref{fig:Halpha-variation}).

The accuracy on the measurement of radial velocities depends on a correct wavelength calibration, as well as the correction of the heliocentric velocity. Also it depends on the method used for obtaining the RV values. The methods are:

\begin{itemize}
\item Cross-correlation method: based on correlating the observed spectrum with a known RVs spectrum, which can be a synthetic, standard or an observed spectrum. 
\item Fitting of line profiles: by fitting the broadened line profile by a Gauss, Voigt or Lorentz distribution. The fitted line center is compared with the observed one. 
\item Mirroring method: comparing the line profile of the observed spectrum with the same line flipped around the laboratory wavelength, and then moving it up to adjust the wings (or core) of the line profile.
\end{itemize}

\begin{figure*}
   \centering
   \includegraphics[width=1\columnwidth]{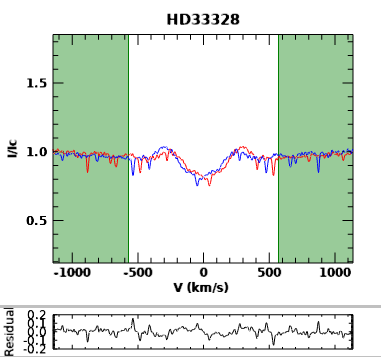}
      \includegraphics[width=1\columnwidth]{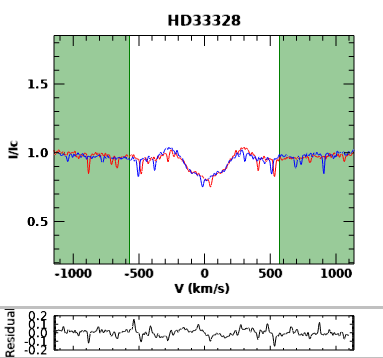}
   \caption{Example of the Mirroring method used on the H$\alpha$ line profile of one spectrum of the Be star HD 33328. \textit{Left:} Observed and flipped spectrum in red and blue, respectively. \textit{Right:} Observed and flipped spectrum shifted 36.80 km s$^{-1}$, corresponding to a RV of 18.40 km s$^{-1}$. The green zone indicates the selection of the continuum. To calculate the $\chi^{2}$ the displayed range in common between the observed and flipped spectrum is used. The residual of the fit is shown below the plots.}
   \label{mirroring}
\end{figure*}

\begin{figure}
\includegraphics[width=1\columnwidth]{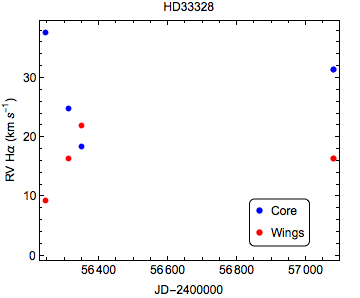}
\caption{Radial velocity (RV) variation of the Be star HD 33328 obtained by using the ``mirroring'' method. } \label{Halpha-variation2} 
\end{figure}

There are advantages and disadvantages for each of these different methods. For example, the first two work well on symmetric line profiles, while the mirroring method is efficient for asymmetric profiles but is strongly subjective. Since Be stars usually present asymmetric line profiles, we used the mirroring method to derive RV values. An example of the implementation in MUFIN of this method is shown in Figure~\ref{mirroring} for the Be star HD 33328 in the H$\alpha$ line profile. The observed and flipped line profiles are represented in red and blue, respectively (left plot). The user can interact with the program to select the continuum section denoted by the green zone in order to derive the error for the $\chi^{2}$ calculation. In the right plot of Figure~\ref{mirroring}, the flipped line profile (blue) is moved to fit the wings of the observed line profile (red). The resulting shift must be divided by 2 taking into account that we only move the flipped line profile. Therefore, the radial velocity value is RV $\approx$ 18.40 km s$^{-1}$, with a $\chi^{2} =$ 2.79 and a continuum average of 0.97 $\pm$ 0.02. The residual of the fit is shown below each plot. We note that the radial velocity measured changes depending on the region to fix (e.g. the core or the wings), since the origin of this movement is attributed to non-radial pulsations of the central star or the presence of a companion. 
We obtained values for both regions, core and wings. In the example (see Fig.~\ref{Halpha-variation2}), radial velocities for the Be star HD 33328 from core and wings are change in the opposite way. Further analysis should be made in order to derive the period of this variation and the origin (applying the method used by \cite{Aerts2010}). More data are needed as well as measurements in Helium and metallic lines, since emission and absorption lines of the Balmer series have different values of the radial velocity. \cite{Merrill1944} studied the Be shell star 48 Librae and they noted that the radial velocity of the Balmer lines in the shell-phase of the star, became more negative with higher series members. This behavior is known as  ``Balmer progression''.
%behaviors at different atomic transitions. \cite{Dachs1990} measured the intensity ratio of Balmer lines in 26 southern early-Be type stars and noted a decrement as move to highest series members, e.g., I(H$\alpha$)/I(H$\beta$) $>$ I(H$\gamma$)/I(H$\beta$), etc.   

Results for all Be stars in the BeSOS website are displayed in the Table~\ref{tab:results} (appendix~\ref{appendix:parte2}). Targets are sorted by HD number. Spectral types (Sp.T) and distances (D), taken from the literature, are displayed in the second and third column, respectively. The parameters obtained by the BeSOS's team are $T_\text{eff}$, $\log$ g, $R_{\star}$ and $v \, \sin i$. The reddening values were also taken from the literature. 

Summarizing, relevant information can be obtained analyzing the changes of the H$\alpha$ line profile. The global intensity variation as well as the DPS variation are a signature of dissipation/formation of a disk. The V/R variation is a signature of an asymmetry and the RV variation will allow us to detect a possible companion for these objects. More variation aspects will be added in the future such as the superposition of the DPS and intensity in order to find correlations (see \citealt{Kanaan2008}).

\subsection{Late B and early A-type stars}
As mentioned before, we constrained the stellar parameters by fitting Tlusty models to the spectroscopic and photometric data where we derive $T_\text{eff}$, $\log$ g, $R_{\star}$ and $v \, \sin i$. For cases with lower temperatures $\leq$ 15000 K, we only give the best SED fit found by fitting Kurucz models. This concerns 16 stars and they are indicated with the superscript letters ``LT'' in the result Table~\ref{tab:results} (Appendix~\ref{appendix:parte2}).
For all these targets, values of the DPS, V/R and radial velocity for H$\alpha$ emission line  are  available on the website. 
 
\subsection{Getting errors on the measurements}
\label{errors}
Every parameter given here was found applying an algorithm that minimizes the chi-square value. Once the parameters for each star were fixed, we calculated the confidence intervals using Monte Carlo simulations. The algorithm starts the minimization with a default value for each free parameter, and then once it converges to a minimal Chi-square reduced value, uses a random value to start again the minimization. The random value is chosen from a fixed range previously defined and to converge the algorithm moves between values within $\sim$ 68$\%$ (1 standard deviation). The user can set the number of times that the algorithm runs, we set the iteration number to 30. From these 30 iterations, the algorithm chooses the lowest and largest Chi-square reduced values and calculates the standard deviation from the central value, i.e, (Value$_\text{Max}$-Value$_\text{Min}$)/2. As this takes a lot of CPU time, we performed this analysis for one random star of each spectral type. Errors are in general very low, in all cases the percentage deviation does not exceed 1.0$\%$ of the original value for $T_\text{eff}$ and $\log$ g, and 2.0$\%$ for $R_{\star}$ and $v\sin i$. The errors are displayed in the results Table (see Table~\ref{tab:results}). It is important to note that our method does not represent an absolute error for the best fit parameters of each target. In any case the interactive interface (MUFIN) is available for BeSOS users by email request. With this, users can obtain a more realistic error value, either by applying the same method or doing it by ``hand''.

\section{Results}  
  
\subsection{Stellar parameters and $v \, \sin i$ distribution}
\begin{figure*}
   \centering
   \includegraphics[width=1\columnwidth]{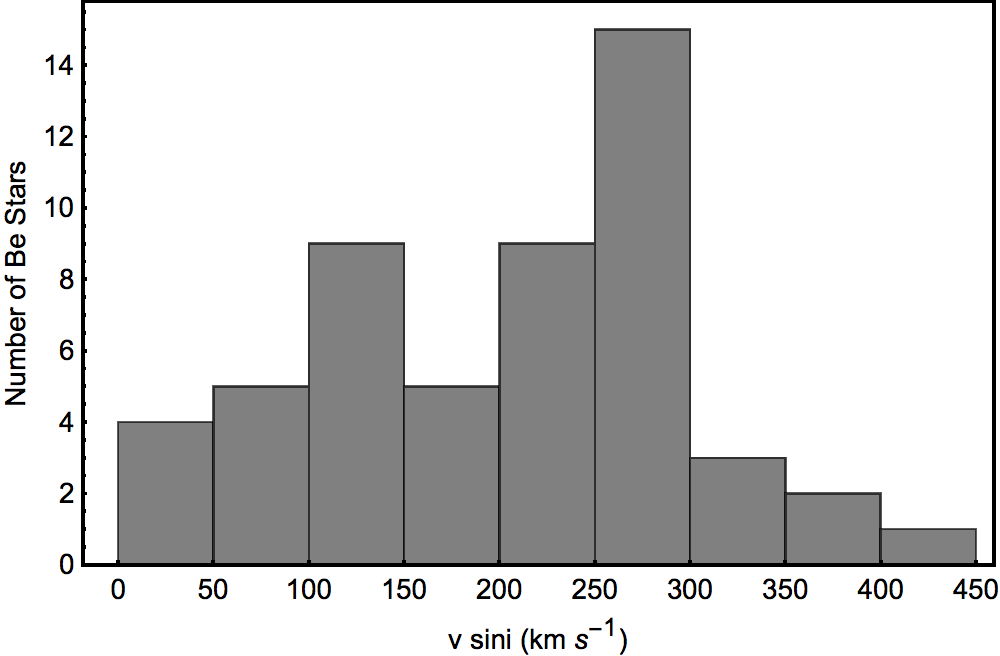}
   \includegraphics[width=1\columnwidth]{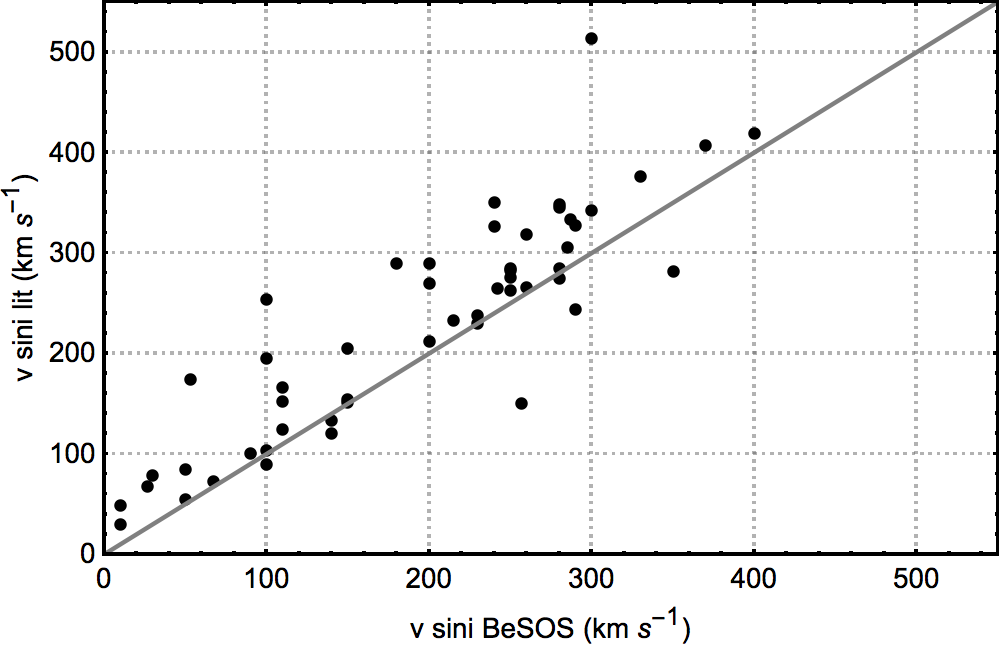}
   \caption{Projected rotational velocities result. \textit{Left:} The distribution of $v \, \sin i$ obtained by the BeSOS team. The distribution peaks between 250 - 300 km s$^{-1}$. \textit{Right:} Comparison between $v \, \sin i$ values obtained by BeSOS and those taken from the literature.}
   \label{vsini}
\end{figure*}

Stellar parameters obtained by fitting the spectral energy distribution as well as the HeI and MgII line profiles of each Be star in BeSOS are shown in the Appendix in Table~\ref{appendix:parte2}. Spectral types and distance values from the literature are reported (from the Simbad database), as well as the stellar parameters, $T_\text{eff}$, $\log g$, $R_{\star}$ and $v \, \sin i$ derived by the BeSOS team. We noted that spectral types derived here are, in general, hotter than those taken from the literature and with luminosity class V (main sequence stars). 
As a first result, we explored the projected rotational velocity distribution of our sample and we found a peak between 250 and 300 km s$^{-1}$ (see left plot of Fig.~\ref{vsini}). A secondary peak is seen between 100 and 150 km s$^{-1}$. Obviously, to search for correlations, we need to deconvolve $v \, \sin i$ to compare how the stellar rotation changes with other stellar parameters, e.g., effective temperature. For the moment, we compared our $v \, \sin i$ values with the literature (see \cite{Fremat2005} and BeSOS website for sources). The result is shown in the right plot of Figure~\ref{vsini} with a diagonal line as reference. A strong correlation is seen between the values, although in general the values of $v \, \sin i$ from the literature are slightly larger than ours.    

\subsection{New emission line stars}
\begin{figure*}
\includegraphics[width=0.7\textheight]{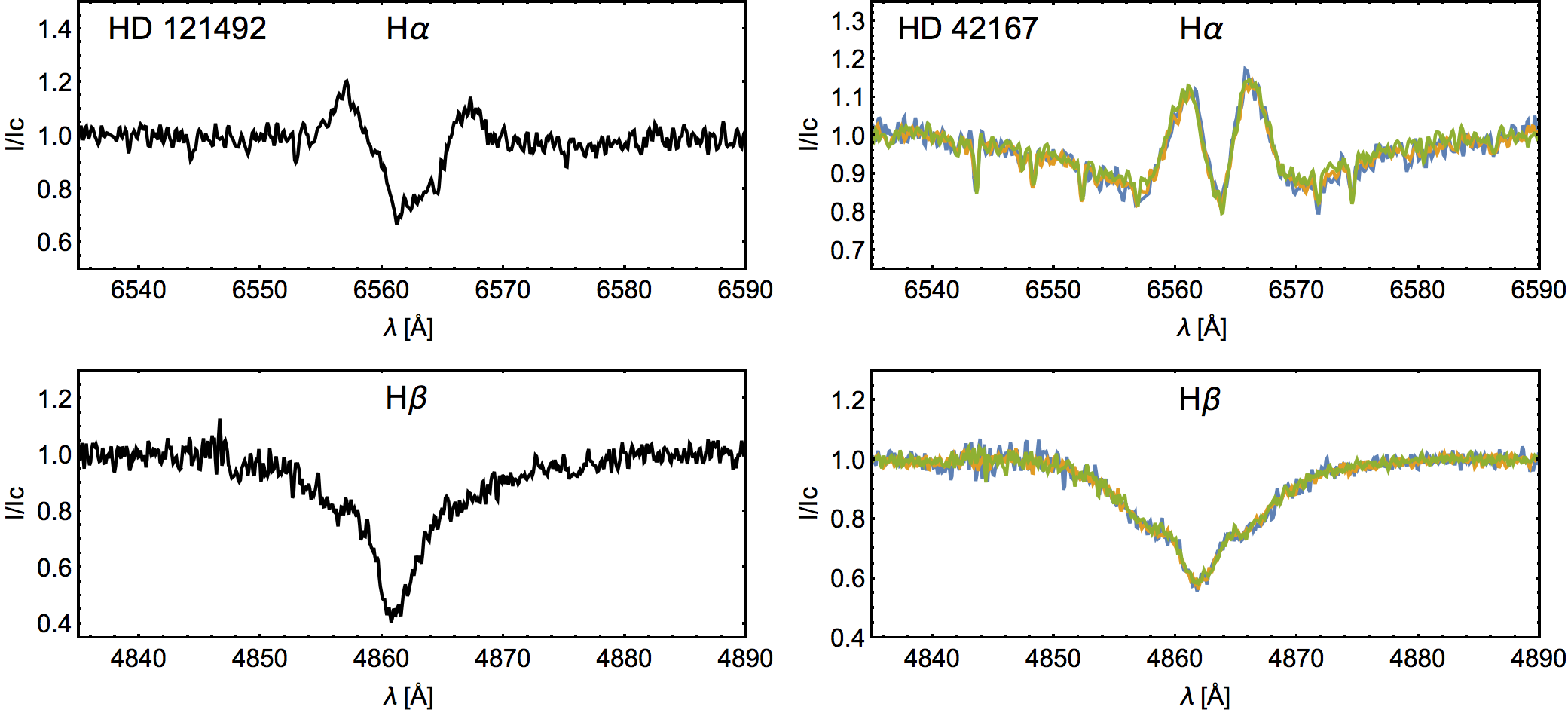}
\caption{H$\alpha$ and H$\beta$ line profiles of the two H$\alpha$ emission stars. Observations were taken using the spectrograph PUCHEROS.} 
\label{NewBestars}
\end{figure*}

The BeSOS survey is dedicated to the study of the variability of Be stars, but also to discover/confirm new Be stars. To date, we have found one classical Be star and one star showing emission in the H$\alpha$ line profile (see Fig.~\ref{NewBestars}). 

These stars are poorly studied and there is no information available in the literature about an existing disk, except for HD 42167 where previous work was done deriving the disk parameters in \cite{Arcos2017}. The spectral type is taken from the literature and is compared with that obtained in this work. 

\begin{figure*}
\includegraphics[width=0.45\textwidth]{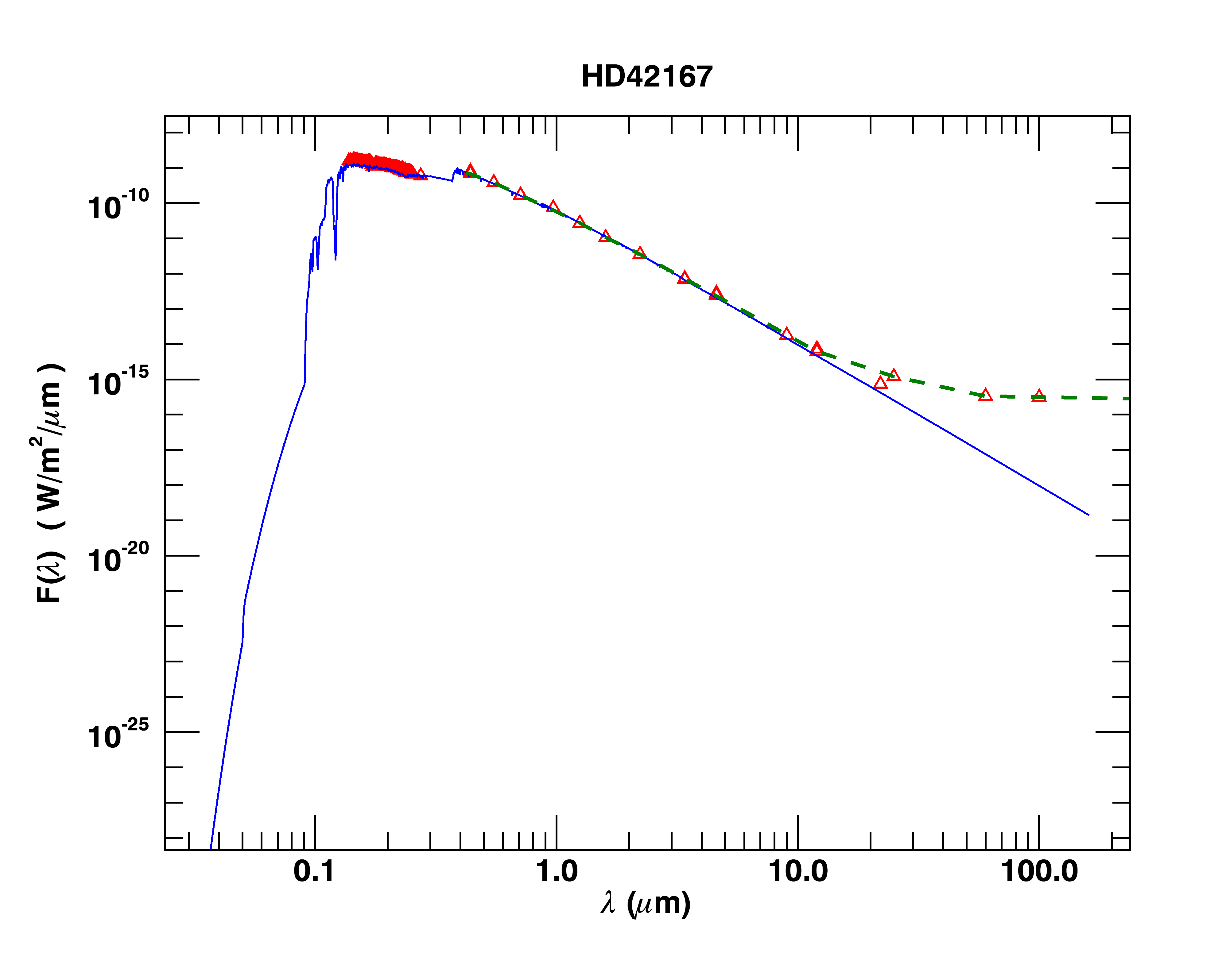}
\includegraphics[width=0.45\textwidth]{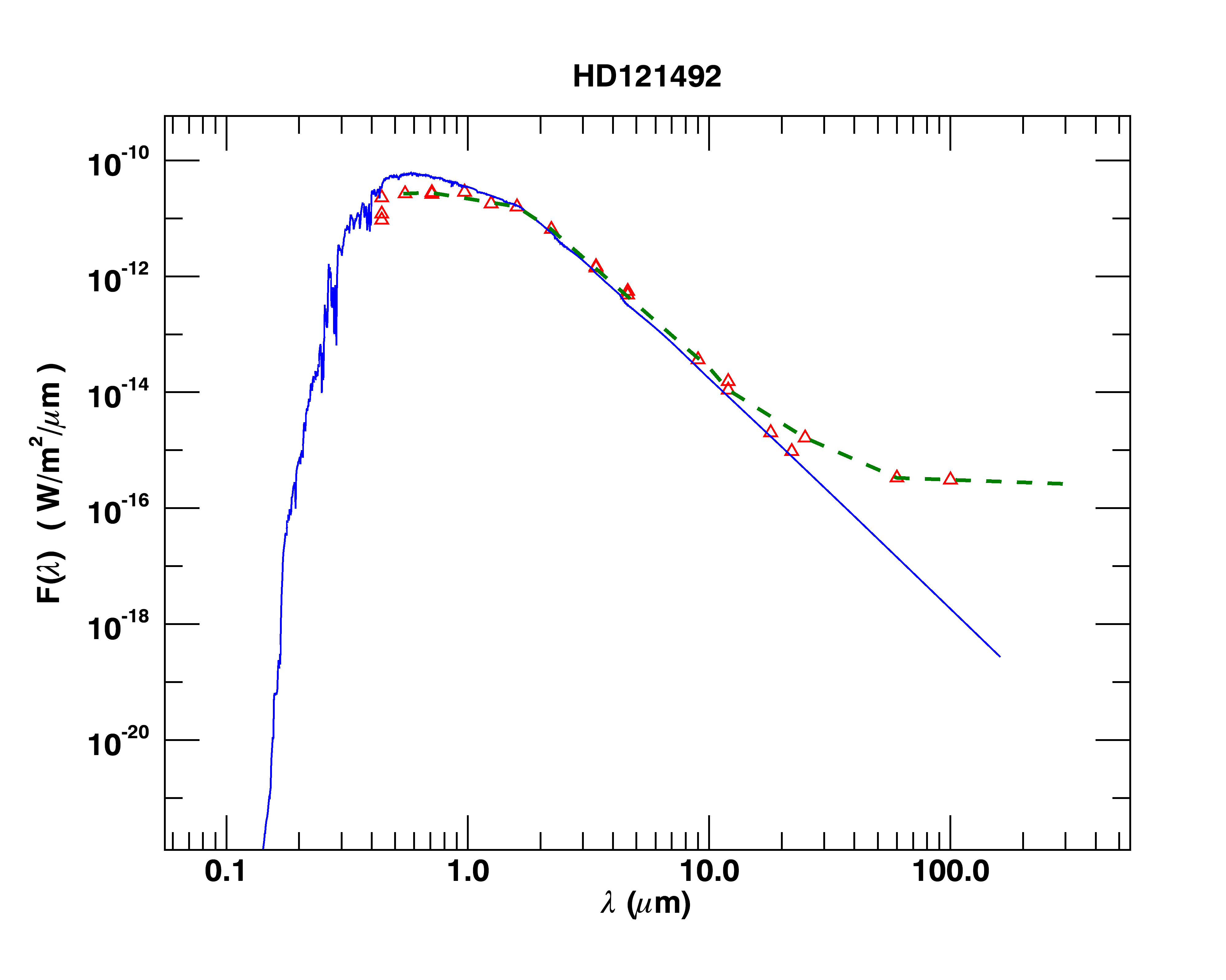}
\caption{Spectral energy distribution of the new classical Be star HD 42167 (left) and the YSO Class III HD 121492 (right). The best Kurucz model, photometric data and IR-excess are represented by a solid blue line, empty red triangles and a  dashed green line. For the YSO object the IR excess starts at $\sim$ 20$\mu$m.} 
\label{sedhd42167}
\end{figure*}

HD 121492 (HIP67988) is a star with a K0 spectral type and with a visual magnitude of 7.88. We have one spectrum taken in Jan 31, 2014 showing a double peaked H$\alpha$ line profile. This object is still not classified, but the SED shows an IR excess in the Far-IR and due to the spectral type the most probable classification is a ``Young Stellar Object'' (YSO) of Class III. The different dust distributions among the YSO classes result in different SEDs. The classification between classes I to III, depend on the IR-excess slope (from 2 up to 20$\mu$m), where Class I is for the stronger and Class III is for the weakest IR-excess, respectively \citep{Andre1993}. The SED is shown in the right plot of Figure~\ref{sedhd42167}. The best Kurucz model is for $T_\text{eff}$ = 4736 $\pm$ 47 K and $\log$ g = 2.00 $\pm$ 0.02 cm s$^{-2}$, with $R_{\star}$ = 18.65 $\pm$ 0.37 $R_{\odot}$ and a distance of 300 $\pm$ 12 pc, which corresponds to a K2III spectral type. 

HD 42167 ($\theta$ Col, HR2177, HIP29034) has a B9IV spectral type with a visual magnitude of 4.99. Using our interactive program to fit the photometric data, the best Kurucz model is for $T_\text{eff}$ = 11430 $\pm$ 114 K and $\log$ g = 3.00 $\pm$ 0.03 cm s$^{-2}$, with $R_{\star}$ = 6.75 $\pm$ 0.14 $R_{\odot}$ and a distance of 221 $\pm$ 9 pc, which corresponds to a B7III spectral type. We have observations from the dates 2014-01-30, 2015-02-25 and 2015-05-06 where the spectra do not show variability during this time. \cite{Royer2002} derived a value of $v \, \sin i \sim$ 250 km s$^{-1} $. The SED is shown in the left plot of the Figure~\ref{sedhd42167}. \\

%HD 62632 (CCDM J07446-1219AB, HIP37750) has an Ap spectral type and is in a double or multiple system. Its visual magnitude is 8.63. We have one spectrum taken in Jan 19, 2013. There is no information available about the $v \, \sin i$ in the literature. The best Kurucz model is for $T_\text{eff}$ = 6400 $\pm$ 64 K and $\log$ g = 4.20 $\pm$ 0.04 cm s$^{-2}$, with $R_{\star}$ = 2.85 $\pm$ 0.06 $R_{\odot}$ and a distance of 348 $\pm$ 14 pc, which corresponds to an F6V spectral type. \\

\begin{table*}
\caption{New stars showing Balmer emission lines}           
\centering                  
\begin{tabular}{c c c c c c c c c c}    
\hline    
\multicolumn{4}{c}{Literature/BeSOS} &\multicolumn{3}{c}{H$\alpha$} & \multicolumn{3}{c}{H$\beta$}\\ 
\hline 
HD & Sp.T & $v \, \sin i$ & Date & EW & V/R  &  DPS   & EW    & V/R  &  DPS  \\   
   &      & (km s$ ^{-1} $) &   & ($\text{\AA}$)&  &  (km s$ ^{-1}$) & ($\text{\AA}$)& &  (km s$ ^{-1} $) \\   
\hline  
121492 & K0/K2III  & --  & Jan 31, 2014  & 1.34  & 1.10 & 464.6  & 4.91  & 0.99  & 510.7 \\
42167 & B9IV/B7III & 250 & Jan 30, 2014  & 2.50  & 0.96 & 229.4  & 4.37  & 1.01  & 305.9  \\
        &      &     & Feb 25, 2015  & 1.97  & 0.95 & 230.6  & 4.26  & 1.02  & 369.6   \\ 
        &      &     & May 06, 2015  & 2.05  & 0.99 & 228.4  & 4.25  & 0.99  & 338.7 \\ 
%62632 & ApSiCr/F6V & --& Jan 19, 2013  & -3.70 & 0.70 & 118.3  & 3.04  & 0.87  & 129.3 \\
\hline                                   %inserts single line
\end{tabular}
\label{tab:newobjects}
\end{table*}

For both objects we calculated V/R, EW and DPS from H$\alpha$ and H$\beta$ line profiles, the results are shown in the Table~\ref{tab:newobjects}.  

\subsection{Physical models fitted on H$\alpha$ emission lines.}
Previous work was carried out by \cite{Arcos2017} to study the observed H$\alpha$ emission lines of 42 Be stars in the BeSOS survey using radiative transfer codes. In that work, the authors used the non-LTE code \textsc{BEDISK} \citep{Sigut2007} and the auxiliary code \textsc{BERAY} \citep{Sigut2011} to solve the transfer equation along a series of rays ($\sim$ 10$^{5}$) to produce theoretical H$\alpha$ line profiles. They took all spectra of each Be star as an independent model to make a statistical study of mass, angular momentum and size of the disk. In total 61 H$\alpha$ emission lines were modeled. The parameters of the best fits are displayed in Table 2 of \cite{Arcos2017}. Moreover, statistical results were based on a percentage of the representative models (best range of models) for each H$\alpha$ emission line, where they found that the disk mass and angular momentum distributions are different between early (B0-B3) and late (B4-B9) subspectral types. Also, they found values of the power-law density distribution (see eq.1 on \cite{Arcos2017}) of  $\langle \log \rho \rangle \sim$ -10.4 to -10.2, and $\langle n \rangle \sim$ 2.0 to 2.5 and disk masses of 3.4$\times$10$^{-9}$ and 9.5$\times$10$^{-10}$ M$_{\star}$ for early and late spectral types, respectively. For the disk size, they calculated the emitting region containing 90$\%$ of the total H$\alpha$ flux given a range of values concentrated between 10 and 15 stellar radius.

\subsection{Complementing other databases}
\begin{figure*}
          \includegraphics[scale=0.65]{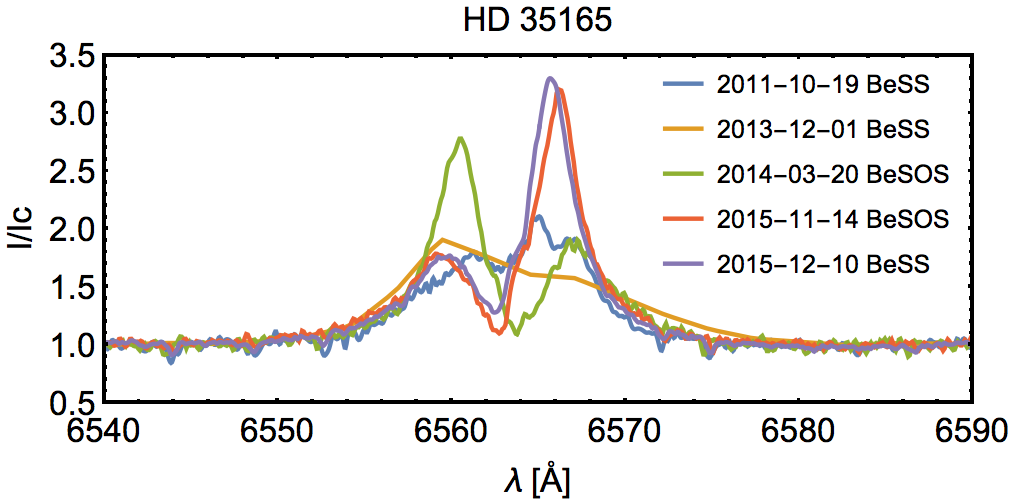}
          \includegraphics[scale=0.65]{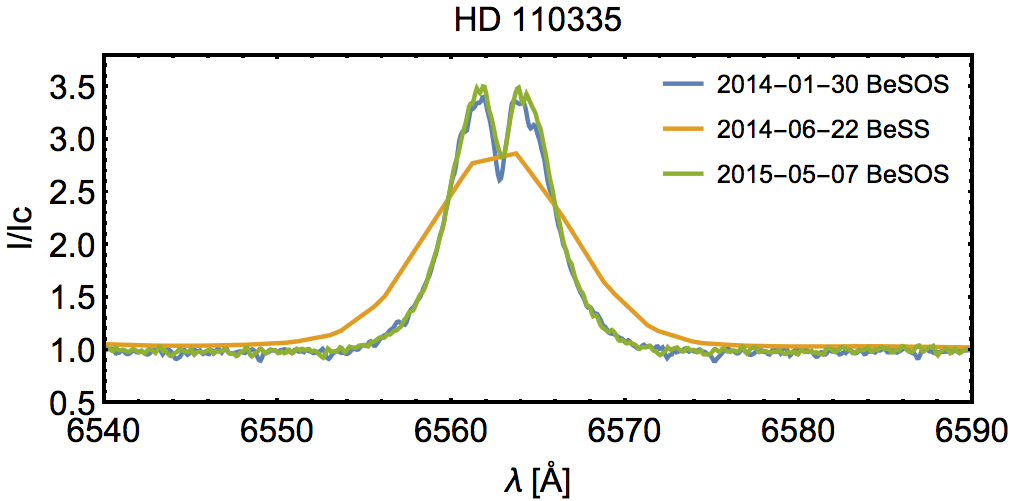}
          \includegraphics[scale=0.65]{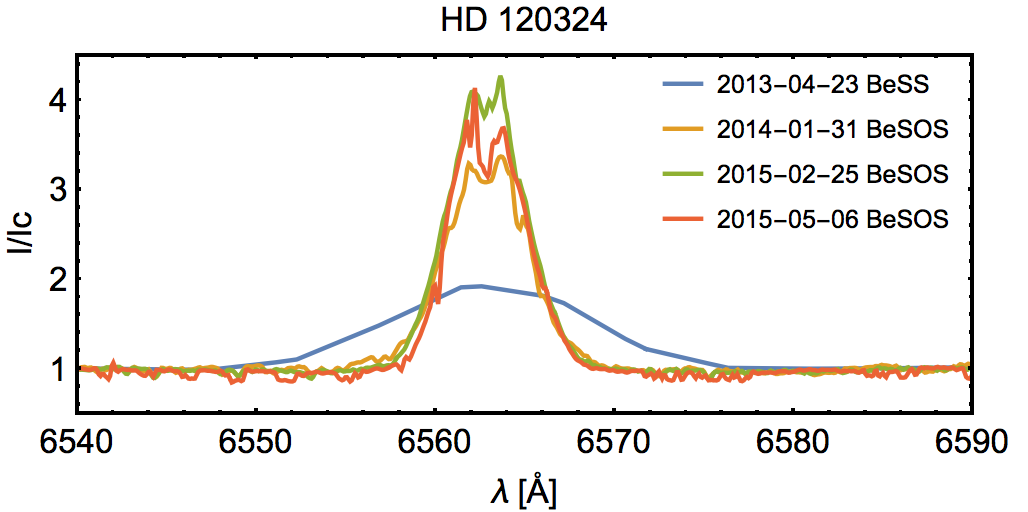}
          \includegraphics[scale=0.65]{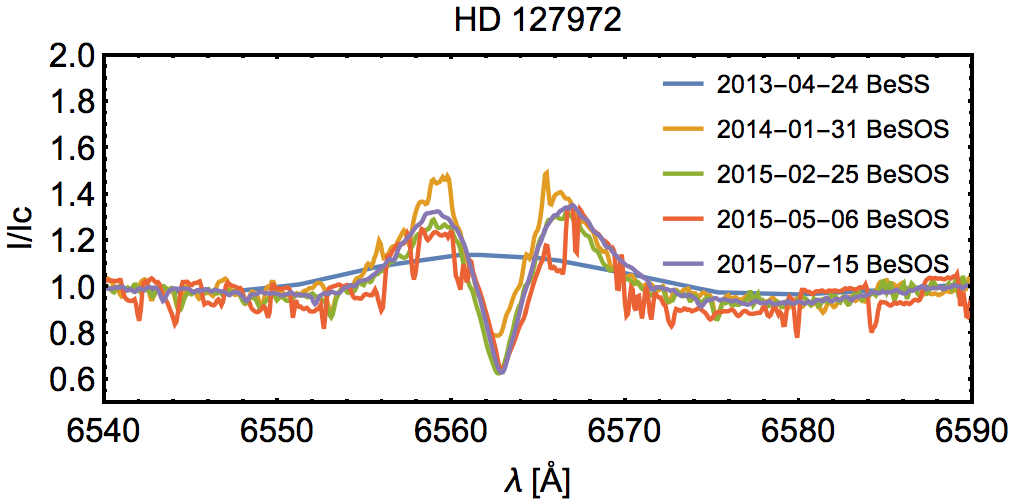}
        \caption{Different kinds of variabilities of four Be stars using BeSOS and BeSS spectra. In the case of HD 35165 a V/R variation is shown. For HD 110335 a stable disk is found. In the case of HD 127972 a shell line profile was discovered by BeSOS as well as a disk dissipation phase. Finally intensity and V/R variation is shown in the case of HD 120324.} \label{comparison} 
  \end{figure*}

BeSOS is not only a survey of southern Be stars, the main goal is to offer to the user complete information about the targets in the survey. 
In the literature, a database containing a large number of observed Be stars in the whole sky is the ``Be Star Spectra'' (BeSS, \url{http://basebe.obspm.fr/basebe/}). BeSS is an open database containing spectra from different instruments and telescopes. The majority of southern Be stars with $V<8$ in BeSS have less than 3 spectra  in the optical (177 out of 281 Be stars). Moreover 49, 59, and 40 Be stars have respectively 0, 1 and 2 spectra. In the BeSOS survey 23 of these Be stars are already observed. BeSOS, in these cases, completes the data from BeSS, in order, for example, to find variation or observe for the first time the target in the optical range. As an example,  we show the star HD 35165 that exhibits a V/R cycle by combining our data with the BeSS data (see Figure~\ref{comparison}). We also show three of BeSOS stars containing only one low resolution optical spectrum in BeSS (HD 110335, HD 120324 and HD 127972). In these examples, BeSOS was used to complete BeSS by adding high resolution spectra (see Figure~\ref{comparison}).
HD 110335 has three different observation epochs, where spectra from BeSOS do not show intensity variation. The flux is almost the same (see EW in the Table~\ref{tab:besosandbess}), but the  H$\alpha$ double-peak is clearly absent from the BeSS spectrum due to the low resolution (e.g. R=1000).
HD 120324 shows a double peaked emission line with intensity and V/R variations. 
In the case of HD 127972, BeSOS allowed us to find evidence for its shell status with a decrease in its intensity pointing to a dissipating disk phase.
Information about these objects: date, observatory site, instrument and resolution are displayed in Table~\ref{tab:besosandbess}, as well as the EW, V/R and DPS, calculated for the H$\alpha$ emission line.

\begin{table*}
\caption{BeSOS vs BeSS objects }
\begin{tabular}{c c c c c c c c c c}
 \hline
HD & Sp.T & $v \, \sin i$ & Date & Observatory & Instrument & Resolution  & EW & V/R & DPS  \\
 &  & (km s$ ^{-1} $) &  & Site &  &  & \text(\AA) &  & (km s$ ^{-1} $) \\    
 \hline
35165 & B2Vnpe & 350.0  & Oct 19, 2011 & Melbourne  & Shelyak eShel & 10000 & -9.4  & 0.98  & 217.0\\
        &        &        & Dec 01, 2013 & Mirranook  & LISA          & 1000  & -11.7 & 1.16  & 240.0\\
        &  B4V   & 240.0  & Mar 20, 2014 & OUC        & PUCHEROS      & 17000 & -11.4 & 1.46  & 294.0\\
        &  B4V   & 240.0  & Nov 14, 2015 & OUC        & PUCHEROS      & 17000 & -12.5 & 0.56  & 308.0\\
        &        &        & Dec 10, 2015 & Shenton Park & LhiresIII   & 18000 & -12.6 & 0.54  & 274.0\\ 
\hline
110335 & B8IV  &        & Jan 30, 2014 & OUC       & PUCHEROS & 17000 & -16.2 & 1.0 & 105.0\\
         & B5III & 208.0  & Jun 22, 2014 & Mirranook & LISA     & 1000  & -19.8 & --  & -- \\
         & B8IV  &        & May 07, 2015 & OUC       & PUCHEROS & 17000 & -16.9 & 1.0 & 112.0\\ 
\hline
120324 & B2Vnpe & 166   & Apr 23, 2013 & Castanet & ALPY     & 600   &  -12.3   & --   & --  \\
         & B1V    & 110.0 & Jan 31, 2014 & OUC      & PUCHEROS & 17000 &  -12.6   & 1.0  & 97.5\\
         & B1V    & 110.0 & Feb 25, 2015 & OUC      & PUCHEROS & 17000 &  -15.3   & 1.0  & 65.7\\
         & B1V    & 110.0 & May 06, 2015 & OUC      & PUCHEROS & 17000 &  -12.0   & 1.0  & 79.9\\
\hline 
127972 & B2Ve & 326.0   & Apr 24, 2013 & Castanet & ALPY  & 600   & -1.8 & --   & -- \\
         & B1V  & 240.0   & Jan 31, 2014 & OUC   & PUCHEROS & 17000 & -1.8 & 1.0  & 308.4\\
         & B1V  & 240.0   & Feb 25, 2015 & OUC   & PUCHEROS & 17000 & -0.1 & 1.0  & 319.8\\
         & B1V  & 240.0   & May 06, 2015 & OUC   & PUCHEROS & 17000 &  0.9 & 0.9  & 376.7\\
         & B1V  & 240.0   & Jul 15, 2015 & OUC   & PUCHEROS & 17000 & -1.0 & 1.0  & 354.9\\     
\hline                                 %inserts single line
\end{tabular}
\label{tab:besosandbess}
\end{table*}

\begin{figure}
\includegraphics[width=0.35\textheight]{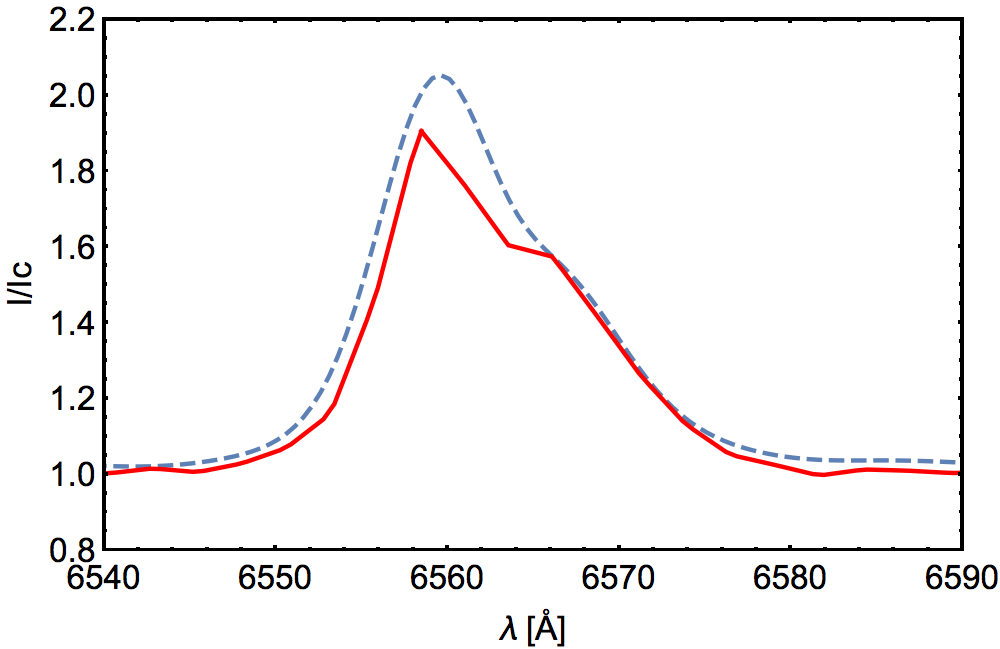}
\caption{Mirrored BeSOS spectrum on Dec, 2015, convoluted with an instrumental resolution of 1000 (dashed blue line) compared with the BeSS spectrum on Dec, 2013 (solid red line). This comparison allowed us to deduce that the maximum intensity of the violet peak occurred on Dec, 2013.} 
\label{fig:conv_HD35165} 
\end{figure}

In the case of HD 35165 observations from various epochs show a clear V/R variation. This kind of variation is associated to a precessing one-armed over density in the disk, and can be modeled using a radiative transfer code \citep{Carciofi2009}.  Spectra from BeSS and BeSOS allowed us to estimate the period of this variation. The maximum intensity of the R peak in the available spectra occurred on Dec, 2015 at I $\sim$ 3.3 (V/R = 0.54). If we consider that date as the maximum intensity for the red peak, then  by symmetry the violet peak should be at the same maximum intensity, when the red peak is at its minimum. We took the spectrum on Dec, 2015 (BeSS database) and we created a mirrored spectrum. Then, we lowered the spectral resolution of this mirrored spectrum (R=18000) to an instrumental resolution of 1000, and we compared it  with the spectrum on Dec, 2013 of the BeSS database. The result is shown in Figure~\ref{fig:conv_HD35165}, where the dashed blue line corresponds to the mirrored spectrum convoluted with R=1000 and the red line corresponds to the spectrum from BeSS. Based on the similarities of both spectra, we can deduce that the maximum of the violet peak (V/R $\sim$ 1.85) occurred around Dec, 2013. 

\begin{figure}
   \centering
   \includegraphics[width=1\columnwidth]{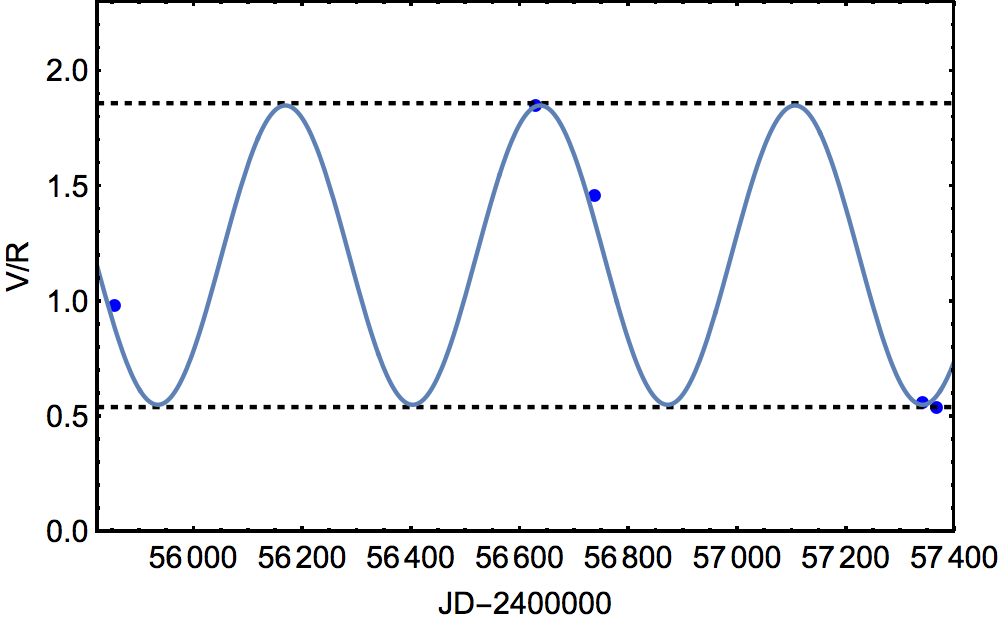}
    \caption{Estimation of the V/R period of HD 35165. Points are taken from the Table~\ref{tab:besosandbess}, where the value on Dec 01, 2013 was replaced by 1.85. A sinusoidal fit is overplot to the data given a period of $\sim$ 480 d. The fit shown here is only for reference, more data are needed to derive a period.}
    \label{fig:PeriodHD35165}
\end{figure}

We do not have enough data to determine a precise period of the V/R cycle variation. By simply fitting a coarse sinusoidal function (assuming V/R $\sim$ 1.85 on Dec, 2013) we estimated a V/R period of $\sim$ 1 year and 4 months ($\sim$ 480 d). Missing points during the 2012 year and future points could give a better period estimation than the derived period here. The period plot is shown in the Figure~\ref{fig:PeriodHD35165}.

\section{Future work}

 \subsection{Interferometry}
We have 13 Be stars in BeSOS observed with the instrument AMBER \citep{Petrov2007}, a near-infrared spectro-interferometric instrument of the VLT interferometer. These observations were carried out during October, 2014 in high-resolution mode (HR = 12000) and centered on the Br$\gamma$ 2.16 $\mu$m line to enable the study of the circumstellar gas kinematics through the Doppler effect. To constrain the geometry and kinematics of the disk, we use a model based on the model proposed by \cite{Delaa2011}. Further explanation and details can be found in that work. The parameters of this model can be classified as follows:
\begin{itemize}
    \item Global geometric parameters: stellar radius ($R_{\star}$), distance ($d$), inclination of the system ($i$) and disk major-axis position angle in the sky (PA).
    \item Global kinematic parameters: rotational velocity ($v_\text{rot}$) at the disk inner radius, expansion velocity at the photosphere ($v_{0}$), terminal velocity ($v_{\infty}$) and exponents of the expansion ($\gamma$) and rotation ($\beta$) velocity power laws.
    \item K-band continuum disk geometry:  contribution of the continuum ($a_\text{c}$) in terms of the Full Width at Half Maximum (FWHM) and the ratio between the flux of the central star and the flux of the envelope ($F_\text{env}$).
    \item Br$\gamma$ disk geometry: extension of the Br$\gamma$ line-emission region in terms of the FWHM and the EW. 
\end{itemize}

The disk is assumed to be in a  Keplerian rotation, therefore the expansion velocity can be negligible and the exponent of the rotation velocity power law $\beta$ takes the value -0.5 \citep{Meilland2012}. Then, the projected rotational velocity is defined as:
\begin{equation}
v_{\phi} = v_\text{rot} \left(\frac{r}{R_{\star}}\right)^{\beta},
\end{equation}
where $v_\text{rot}$ is considered proportional to the critical velocity:
\begin{equation}
v_\text{c} = \sqrt{\frac{2 G M_{\star}}{3 R_\text{p}}},  
\end{equation}
with $R_\text{p}$ the polar radius and $r$ the distance to the center of the star. \\

\begin{figure*}
\includegraphics[width=.5\textheight]{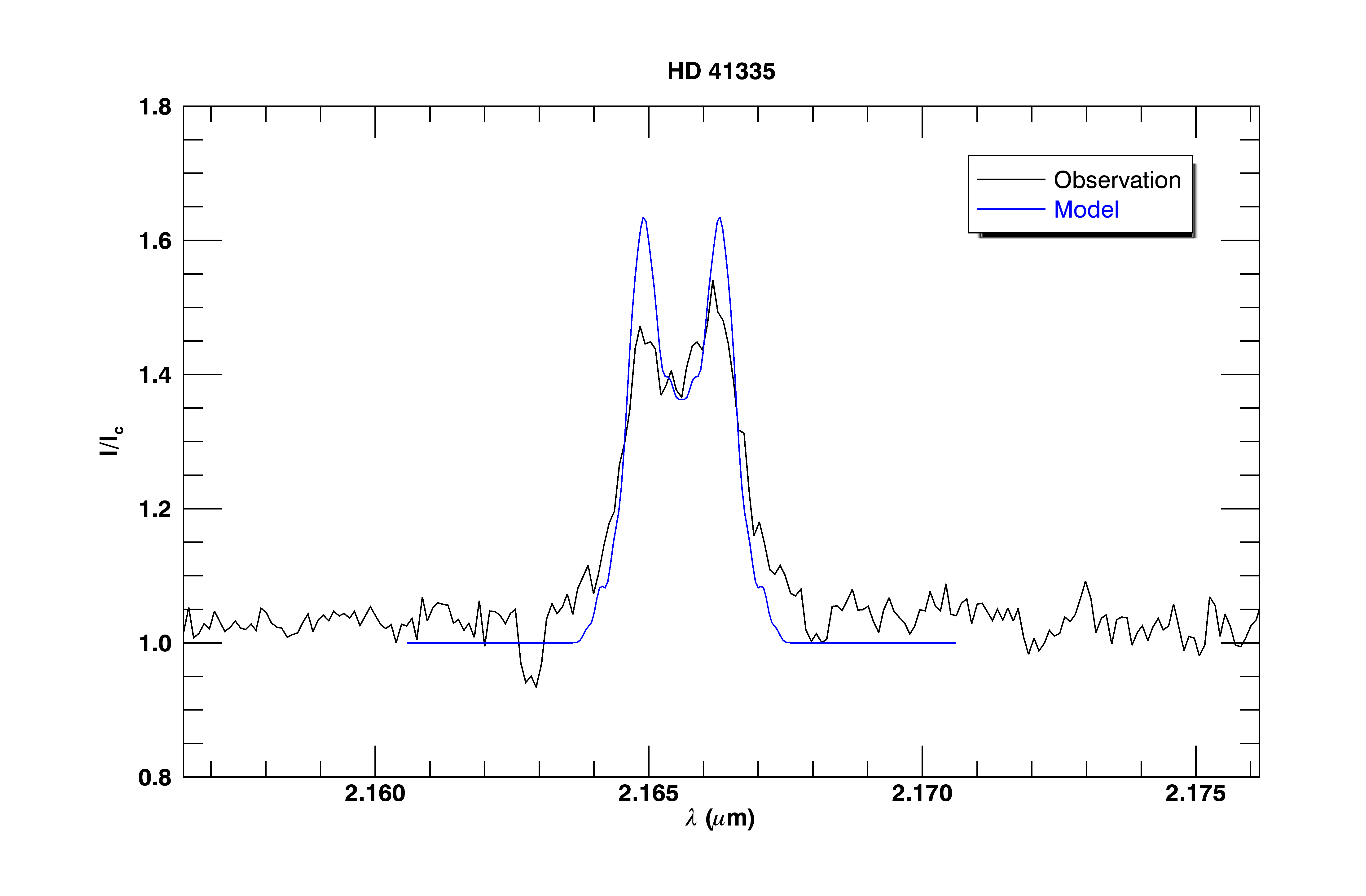}
\includegraphics[width=.7\textheight]{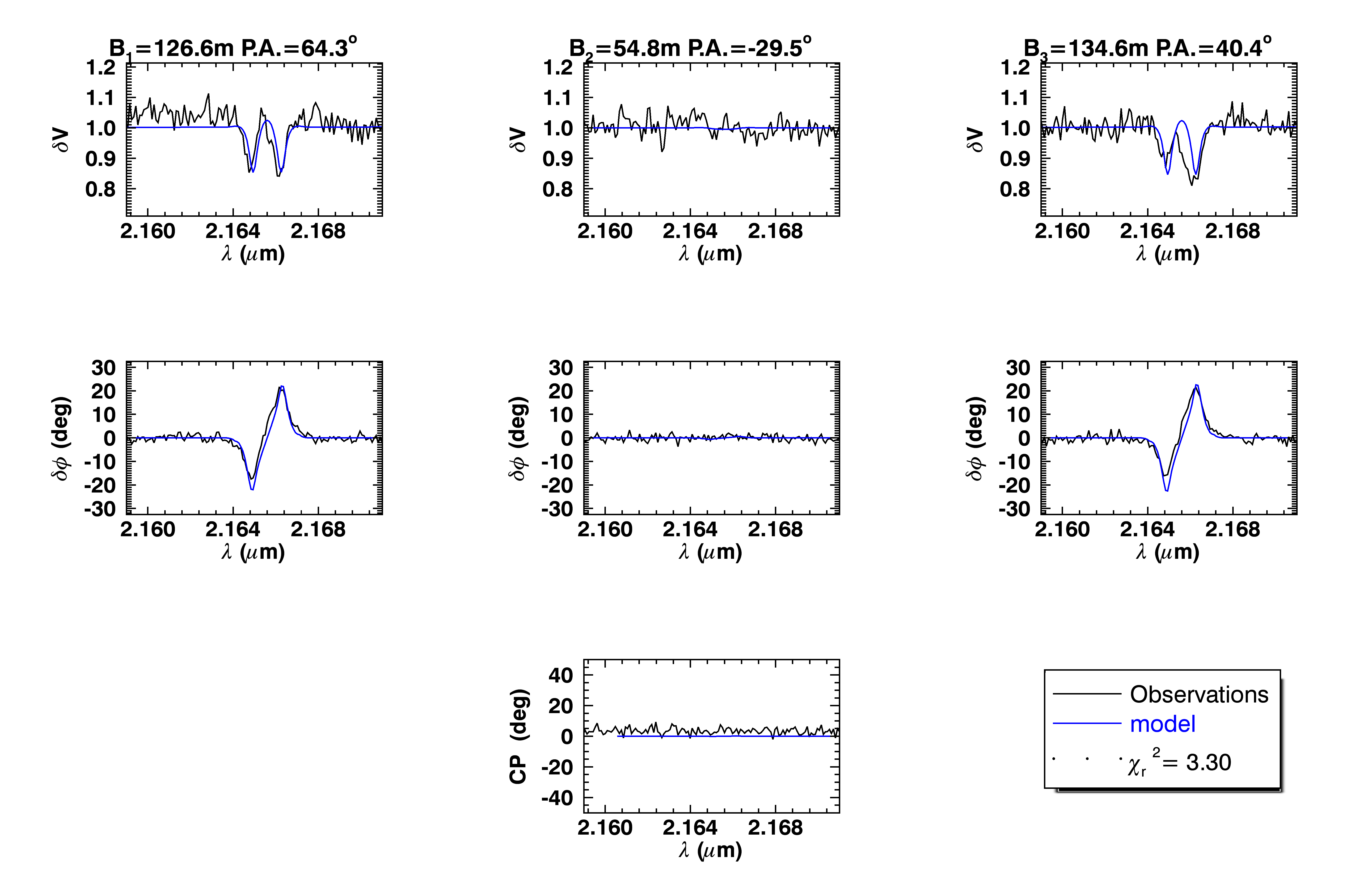}
\caption{Best fit for the interferometric data of the Be star HD 41335. The assumed model is explained in detail in \citep{Delaa2011}. This is only an example of what BeSOS will offer in the interferometry section for 13 Be stars.} \label{ex_int1}
\end{figure*}

We are working on constraining the stellar and disk parameters of these 13 Be stars by using the optical and infrared spectra, as well as the interferometric observables: differential visibility ($\delta$V), differential phase ($\delta \phi$) and closure phase (CP). 
As an example of the model described above, we show the best fit found for the Be star \hbox{HD 41335}. In this case we have one interferometric measurement with AMBER (3 baselines), where the double-peaked line profile, visibility and phase variations present small asymmetries.  We set the stellar radius, the distance and the projected rotational velocity to those obtained in the photometric and spectroscopic fit modelling (see Table~\ref{tab:results}). The EW and the $F_\text{env}$ were measured from the Br$\gamma$ emission line and the SED, respectively. The values are $F_\text{env}$ = 71$\%$ and EW(Br$\gamma$) = -12.11 $\text{\AA}$. The best fit was found for an inclination angle of i = 70 $\pm$ 2 $^{\circ}$ (giving a rotational velocity $\sim$ 350 $\pm$ 5 km s$^{-1}$) and a PA = 57$^{\circ}$. The fit is relatively good with a $\chi^{2}$ value of 3.30. The result is shown in the Figure~\ref{ex_int1}, where the upper panel shows the fit on the emission line (observation in black and model in blue). The lower panel shows the fit of the differential visibility and phase for each baseline (the observation in black and the model in blue). The baselines and position angles are shown above each column plot.

\subsection{Supergiant stars}
The BeSOS website contains 17 OBA supergiant stars, which were observed to be included as a part of a PhD thesis work on stellar winds \citep[See e.g.,][]{Haucke2012,Arcos2014,Haucke2015}. These stars are observed with the same conditions as the Be stars. The spectra are available for the user, in addition to information from the literature. Measurements on H$\alpha$ line profiles will be given over time as these stars show variability in the line because they are losing mass through their strong stellar winds \citep{CAK1975}.

\section{Conclusions}
BeSOS is a tool that offers to the Be and massive star community a consistent survey of spectra that the user can interactively explore and download. BeSOS contains stellar parameters compiled from the literature, as well as stellar and circumstellar parameters calculated by us based on the spectra, photometric and interferometric data. While BeSOS is an ongoing work, mainly on the analysis part, the user already has access to the data and available analysis at \url{http://besos.ifa.uv.cl}.

BeSOS will also allow us to increase the number of Be stars observed and the number of spectra, as well as confirm and discover new Be stars, allowing us to increase the sample for a statistical study. It is important to note that a full study of $v \, \sin i$ is being carried out by our team based on the BeSOS survey using the Tlusty code \citep{Hubeny1995} and our interferometric observations, in order to derive the inclination angle. This study will be published in BeSOS.\\

The survey allowed us to discover two new emission lines of which one is a classical Be star (e.g. HD 42167 ).\\

Finally, we would appreciate if our names (S. Kanaan, C. Arcos $\&$ L. Vanzi) are considered as co-authors in any work published that makes use of BeSOS data.

\section*{Acknowledgements}
The authors would like to thank the anonymous referee for suggestions and comments that helped improve this paper, and also to Graeme Candlish for correcting the spelling of this work. S.K. and C.A. acknowledge partial support from FONDECYT Iniciacion throught grant 11130702, CONICYT Capital Humano Avanzado project N 7912010046, Gemini-CONICYT 32120033, PUC observatory for the telescope time used to obtain the spectra and the Centro de Astrof\'isica de Valpara\'iso. C.A and I.A acknowledge Fondo Institucional de Becas FIB-UV. C.A thanks to BECAS DE DOCTORADO NACIONAL CONICYT 2016-2017. L.V. acknowledges support from FONDECYT N 1171364. We also acknowledge Patricio Ramirez and Felipe Sobarzo from the Universidad T\'ecnica Federico Santa Mar\'ia, Valpara\'iso, Chile for developing the BeSOS website. This work has made use of: the BeSS database, operated at LESIA, Observatoire de Meudon, France: \url{http://basebe.obspm.fr}, the SIMBAD database, operated at CDS, Strasbourg, France and of the VizieR catalogue access tool, CDS, Strasbourg, France. 

%%%%%%%%%%%%%%%%%%%%%%%%%%%%%%%%%%%%%%%%%%%%%%%%%%



%\bibliographystyle{mnras}

\bibliographystyle{mnras}
\bibliography{References}

\begin{thebibliography}{}
\makeatletter
\relax
\def\mn@urlcharsother{\let\do\@makeother \do\$\do\&\do\#\do\^\do\_\do\%\do\~}
\def\mn@doi{\begingroup\mn@urlcharsother \@ifnextchar [ {\mn@doi@}
  {\mn@doi@[]}}
\def\mn@doi@[#1]#2{\def\@tempa{#1}\ifx\@tempa\@empty \href
  {http://dx.doi.org/#2} {doi:#2}\else \href {http://dx.doi.org/#2} {#1}\fi
  \endgroup}
\def\mn@eprint#1#2{\mn@eprint@#1:#2::\@nil}
\def\mn@eprint@arXiv#1{\href {http://arxiv.org/abs/#1} {{\tt arXiv:#1}}}
\def\mn@eprint@dblp#1{\href {http://dblp.uni-trier.de/rec/bibtex/#1.xml}
  {dblp:#1}}
\def\mn@eprint@#1:#2:#3:#4\@nil{\def\@tempa {#1}\def\@tempb {#2}\def\@tempc
  {#3}\ifx \@tempc \@empty \let \@tempc \@tempb \let \@tempb \@tempa \fi \ifx
  \@tempb \@empty \def\@tempb {arXiv}\fi \@ifundefined
  {mn@eprint@\@tempb}{\@tempb:\@tempc}{\expandafter \expandafter \csname
  mn@eprint@\@tempb\endcsname \expandafter{\@tempc}}}

\bibitem[\protect\citeauthoryear{{Aerts} et~al.,}{{Aerts}
  et~al.}{2010}]{Aerts2010}
{Aerts} C.,  et~al., 2010, \mn@doi [\aap] {10.1051/0004-6361/201014124}, \href
  {http://adsabs.harvard.edu/abs/2010A%26A...513L..11A} {513, L11}

\bibitem[\protect\citeauthoryear{{Andre}, {Ward-Thompson}  \&
  {Barsony}}{{Andre} et~al.}{1993}]{Andre1993}
{Andre} P.,  {Ward-Thompson} D.,   {Barsony} M.,  1993, \mn@doi [\apj]
  {10.1086/172425}, \href {http://adsabs.harvard.edu/abs/1993ApJ...406..122A}
  {406, 122}

\bibitem[\protect\citeauthoryear{{Arcos}, {Cur{\'e}}, {Kanaan}, {Cidale}  \&
  {Haucke}}{{Arcos} et~al.}{2014}]{Arcos2014}
{Arcos} C.,  {Cur{\'e}} M.,  {Kanaan} S.,  {Cidale} L.~S.,   {Haucke} M.,
  2014, in Revista Mexicana de Astronomia y Astrofisica Conference Series. pp
  147--147

\bibitem[\protect\citeauthoryear{{Arcos}, {Jones}, {Sigut}, {Kanaan}  \&
  {Cur{\'e}}}{{Arcos} et~al.}{2017}]{Arcos2017}
{Arcos} C.,  {Jones} C.~E.,  {Sigut} T.~A.~A.,  {Kanaan} S.,   {Cur{\'e}} M.,
  2017, \mn@doi [\apj] {10.3847/1538-4357/aa6f5f}, \href
  {http://adsabs.harvard.edu/abs/2017ApJ...842...48A} {842, 48}

\bibitem[\protect\citeauthoryear{{Brahm}, {Jord{\'a}n}  \& {Espinoza}}{{Brahm}
  et~al.}{2017}]{Brahm2017}
{Brahm} R.,  {Jord{\'a}n} A.,   {Espinoza} N.,  2017, \mn@doi [\pasp]
  {10.1088/1538-3873/aa5455}, \href
  {http://adsabs.harvard.edu/abs/2017PASP..129c4002B} {129, 034002}

\bibitem[\protect\citeauthoryear{{Carciofi}}{{Carciofi}}{2001}]{Carciofi2001}
{Carciofi} A.~C.,  2001, PhD thesis, Instituto de Astronomia, Geof{\'{\i}}sica
  e Ci{\^e}ncias Atmosf{\'e}ricas, Universidade de S{\~a}o Paulo, Brazil

\bibitem[\protect\citeauthoryear{{Carciofi}, {Okazaki}, {Le Bouquin}, {{\v
  S}tefl}, {Rivinius}, {Baade}, {Bjorkman}  \& {Hummel}}{{Carciofi}
  et~al.}{2009}]{Carciofi2009}
{Carciofi} A.~C.,  {Okazaki} A.~T.,  {Le Bouquin} J.-B.,  {{\v S}tefl} S.,
  {Rivinius} T.,  {Baade} D.,  {Bjorkman} J.~E.,   {Hummel} C.~A.,  2009,
  \mn@doi [\aap] {10.1051/0004-6361/200810962}, \href
  {http://adsabs.harvard.edu/abs/2009A%26A...504..915C} {504, 915}

\bibitem[\protect\citeauthoryear{{Castor}, {Abbott}  \& {Klein}}{{Castor}
  et~al.}{1975}]{CAK1975}
{Castor} J.~I.,  {Abbott} D.~C.,   {Klein} R.~I.,  1975, \mn@doi [\apj]
  {10.1086/153315}, \href {http://adsabs.harvard.edu/abs/1975ApJ...195..157C}
  {195, 157}

\bibitem[\protect\citeauthoryear{{Collins}}{{Collins}}{1974}]{Collins1974}
{Collins} II G.~W.,  1974, \mn@doi [\apj] {10.1086/152950}, \href
  {http://adsabs.harvard.edu/abs/1974ApJ...191..157C} {191, 157}

\bibitem[\protect\citeauthoryear{{Dachs}, {Rohe}  \& {Loose}}{{Dachs}
  et~al.}{1990}]{Dachs1990}
{Dachs} J.,  {Rohe} D.,   {Loose} A.~S.,  1990, \aap, \href
  {http://adsabs.harvard.edu/abs/1990A%26A...238..227D} {238, 227}

\bibitem[\protect\citeauthoryear{{Delaa} et~al.,}{{Delaa}
  et~al.}{2011}]{Delaa2011}
{Delaa} O.,  et~al., 2011, \mn@doi [\aap] {10.1051/0004-6361/201015639}, \href
  {http://adsabs.harvard.edu/abs/2011A%26A...529A..87D} {529, A87}

\bibitem[\protect\citeauthoryear{{Domiciano de Souza} et~al.,}{{Domiciano de
  Souza} et~al.}{2014}]{Domiciano2014}
{Domiciano de Souza} A.,  et~al., 2014, \mn@doi [\aap]
  {10.1051/0004-6361/201424144}, \href
  {http://adsabs.harvard.edu/abs/2014A%26A...569A..10D} {569, A10}

\bibitem[\protect\citeauthoryear{{Fr{\'e}mat}, {Zorec}, {Hubert}  \&
  {Floquet}}{{Fr{\'e}mat} et~al.}{2005}]{Fremat2005}
{Fr{\'e}mat} Y.,  {Zorec} J.,  {Hubert} A.-M.,   {Floquet} M.,  2005, \mn@doi
  [\aap] {10.1051/0004-6361:20042229}, \href
  {http://adsabs.harvard.edu/abs/2005A%26A...440..305F} {440, 305}

\bibitem[\protect\citeauthoryear{{Hanuschik}}{{Hanuschik}}{1996}]{Hanuschik1996a}
{Hanuschik} R.~W.,  1996, \aap, \href
  {http://adsabs.harvard.edu/abs/1996A%26A...308..170H} {308, 170}

\bibitem[\protect\citeauthoryear{{Haucke}, {Cidale}, {Venero}, {Cochetti},
  {Torres}, {Arcos}  \& {Cur{\'e}}}{{Haucke} et~al.}{2012}]{Haucke2012}
{Haucke} M.,  {Cidale} L.~S.,  {Venero} R.~O.~J.,  {Cochetti} Y.,  {Torres} A.,
   {Arcos} C.,   {Cur{\'e}} M.,  2012, Boletin de la Asociacion Argentina de
  Astronomia La Plata Argentina, \href
  {http://adsabs.harvard.edu/abs/2012BAAA...55...95H} {55, 95}

\bibitem[\protect\citeauthoryear{{Haucke}, {Araya}, {Arcos}, {Cur{\'e}},
  {Cidale}, {Kanaan}, {Venero}  \& {Kraus}}{{Haucke} et~al.}{2015}]{Haucke2015}
{Haucke} M.,  {Araya} I.,  {Arcos} C.,  {Cur{\'e}} M.,  {Cidale} L.,  {Kanaan}
  S.,  {Venero} R.,   {Kraus} M.,  2015, in {Meynet} G.,  {Georgy} C.,  {Groh}
  J.,   {Stee} P.,  eds,  IAU Symposium Vol. 307, New Windows on Massive Stars.
  pp 104--105, \mn@doi{10.1017/S1743921314006401}

\bibitem[\protect\citeauthoryear{{Hubeny} \& {Lanz}}{{Hubeny} \&
  {Lanz}}{1995}]{Hubeny1995}
{Hubeny} I.,  {Lanz} T.,  1995, \mn@doi [\apj] {10.1086/175226}, \href
  {http://adsabs.harvard.edu/abs/1995ApJ...439..875H} {439, 875}

\bibitem[\protect\citeauthoryear{{Infante}, {Guesalaga}, {Vanzi}, {Jord{\'a}n},
  {Padilla}  \& {Guzm{\'a}n}}{{Infante} et~al.}{2010}]{Infante2010}
{Infante} L.,  {Guesalaga} A.,  {Vanzi} L.,  {Jord{\'a}n} A.,  {Padilla} N.,
  {Guzm{\'a}n} D.,  2010, in 25th Texas Symposium on Relativistic Astrophysics.
  p.~250

\bibitem[\protect\citeauthoryear{{Kanaan}, {Meilland}, {Stee}, {Zorec},
  {Domiciano de Souza}, {Fr{\'e}mat}  \& {Briot}}{{Kanaan}
  et~al.}{2008}]{Kanaan2008}
{Kanaan} S.,  {Meilland} A.,  {Stee} P.,  {Zorec} J.,  {Domiciano de Souza} A.,
   {Fr{\'e}mat} Y.,   {Briot} D.,  2008, \mn@doi [\aap]
  {10.1051/0004-6361:20078868}, \href
  {http://adsabs.harvard.edu/abs/2008A%26A...486..785K} {486, 785}

\bibitem[\protect\citeauthoryear{{Kurucz}}{{Kurucz}}{1979}]{kurucz1979}
{Kurucz} R.~L.,  1979, \mn@doi [\apjs] {10.1086/190589}, \href
  {http://adsabs.harvard.edu/abs/1979ApJS...40....1K} {40, 1}

\bibitem[\protect\citeauthoryear{{Lee}, {Osaki}  \& {Saio}}{{Lee}
  et~al.}{1991}]{Lee1991}
{Lee} U.,  {Osaki} Y.,   {Saio} H.,  1991, \mn@doi [\mnras]
  {10.1093/mnras/250.2.432}, \href
  {http://adsabs.harvard.edu/abs/1991MNRAS.250..432L} {250, 432}

\bibitem[\protect\citeauthoryear{{Meilland} et~al.,}{{Meilland}
  et~al.}{2007}]{Meilland2007}
{Meilland} A.,  et~al., 2007, \mn@doi [\aap] {10.1051/0004-6361:20064848},
  \href {http://adsabs.harvard.edu/abs/2007A%26A...464...59M} {464, 59}

\bibitem[\protect\citeauthoryear{{Meilland}, {Millour}, {Kanaan}, {Stee},
  {Petrov}, {Hofmann}, {Natta}  \& {Perraut}}{{Meilland}
  et~al.}{2012}]{Meilland2012}
{Meilland} A.,  {Millour} F.,  {Kanaan} S.,  {Stee} P.,  {Petrov} R.,
  {Hofmann} K.-H.,  {Natta} A.,   {Perraut} K.,  2012, \mn@doi [\aap]
  {10.1051/0004-6361/201117955}, \href
  {http://adsabs.harvard.edu/abs/2012A%26A...538A.110M} {538, A110}

\bibitem[\protect\citeauthoryear{{Merrill} \& {Sanford}}{{Merrill} \&
  {Sanford}}{1944}]{Merrill1944}
{Merrill} P.~W.,  {Sanford} R.~F.,  1944, \mn@doi [\apj] {10.1086/144633},
  \href {http://adsabs.harvard.edu/abs/1944ApJ...100...14M} {100, 14}

\bibitem[\protect\citeauthoryear{{Neiner}, {de Batz}, {Cochard}, {Floquet},
  {Mekkas}  \& {Desnoux}}{{Neiner} et~al.}{2011}]{Neiner2011}
{Neiner} C.,  {de Batz} B.,  {Cochard} F.,  {Floquet} M.,  {Mekkas} A.,
  {Desnoux} V.,  2011, \mn@doi [\aj] {10.1088/0004-6256/142/5/149}, \href
  {http://cdsads.u-strasbg.fr/abs/2011AJ....142..149N} {142, 149}

\bibitem[\protect\citeauthoryear{{Ochsenbein}, {Bauer}  \&
  {Marcout}}{{Ochsenbein} et~al.}{2000}]{Vizier2000}
{Ochsenbein} F.,  {Bauer} P.,   {Marcout} J.,  2000, \mn@doi [\aaps]
  {10.1051/aas:2000169}, \href
  {http://cdsads.u-strasbg.fr/abs/2000A%26AS..143...23O} {143, 23}

\bibitem[\protect\citeauthoryear{{Petrov} et~al.,}{{Petrov}
  et~al.}{2007}]{Petrov2007}
{Petrov} R.~G.,  et~al., 2007, \mn@doi [\aap] {10.1051/0004-6361:20066496},
  \href {http://adsabs.harvard.edu/abs/2007A%26A...464....1P} {464, 1}

\bibitem[\protect\citeauthoryear{{Porter} \& {Rivinius}}{{Porter} \&
  {Rivinius}}{2003}]{Porter2003}
{Porter} J.~M.,  {Rivinius} T.,  2003, \mn@doi [\pasp] {10.1086/378307}, \href
  {http://adsabs.harvard.edu/abs/2003PASP..115.1153P} {115, 1153}

\bibitem[\protect\citeauthoryear{{Rivinius}, {Carciofi}  \&
  {Martayan}}{{Rivinius} et~al.}{2013}]{Rivinius2013}
{Rivinius} T.,  {Carciofi} A.~C.,   {Martayan} C.,  2013, \mn@doi [\aapr]
  {10.1007/s00159-013-0069-0}, \href
  {http://adsabs.harvard.edu/abs/2013A%26ARv..21...69R} {21, 69}

\bibitem[\protect\citeauthoryear{{Royer}, {Grenier}, {Baylac}, {Gomez}  \&
  {Zorec}}{{Royer} et~al.}{2002}]{Royer2002}
{Royer} F.,  {Grenier} S.,  {Baylac} M.-O.,  {Gomez} A.~E.,   {Zorec} J.,
  2002, VizieR Online Data Catalog, \href
  {http://adsabs.harvard.edu/abs/2002yCat..33930897R} {339, 30897}

\bibitem[\protect\citeauthoryear{{Schultz} \& {Wiemer}}{{Schultz} \&
  {Wiemer}}{1975}]{Schultz1975}
{Schultz} G.~V.,  {Wiemer} W.,  1975, \aap, \href
  {http://adsabs.harvard.edu/abs/1975A%26A....43..133S} {43, 133}

\bibitem[\protect\citeauthoryear{{Sigut}}{{Sigut}}{2011}]{Sigut2011}
{Sigut} T.~A.~A.,  2011, in {Neiner} C.,  {Wade} G.,  {Meynet} G.,   {Peters}
  G.,  eds,  IAU Symposium Vol. 272, Active OB Stars: Structure, Evolution,
  Mass Loss, and Critical Limits. pp 426--427,
  \mn@doi{10.1017/S1743921311011045}

\bibitem[\protect\citeauthoryear{{Sigut} \& {Jones}}{{Sigut} \&
  {Jones}}{2007}]{Sigut2007}
{Sigut} T.~A.~A.,  {Jones} C.~E.,  2007, \mn@doi [\apj] {10.1086/521209}, \href
  {http://adsabs.harvard.edu/abs/2007ApJ...668..481S} {668, 481}

\bibitem[\protect\citeauthoryear{{Silaj} et~al.,}{{Silaj}
  et~al.}{2016}]{Silaj2016}
{Silaj} J.,  et~al., 2016, \mn@doi [\apj] {10.3847/0004-637X/826/1/81}, \href
  {http://adsabs.harvard.edu/abs/2016ApJ...826...81S} {826, 81}

\bibitem[\protect\citeauthoryear{{Tody}}{{Tody}}{1993}]{Tody1993}
{Tody} D.,  1993, in {Hanisch} R.~J.,  {Brissenden} R.~J.~V.,   {Barnes} J.,
  eds,  Astronomical Society of the Pacific Conference Series Vol. 52,
  Astronomical Data Analysis Software and Systems II. San Francisco, Calif. :
  Astronomical Society of the Pacific, p.~173

\bibitem[\protect\citeauthoryear{{Vanzi} et~al.,}{{Vanzi}
  et~al.}{2012}]{Vanzi2012}
{Vanzi} L.,  et~al., 2012, \mn@doi [\mnras] {10.1111/j.1365-2966.2012.21382.x},
  \href {http://adsabs.harvard.edu/abs/2012MNRAS.424.2770V} {424, 2770}

\bibitem[\protect\citeauthoryear{{Vogt}, {Barrera}  \& {Navarro}}{{Vogt}
  et~al.}{1990}]{Vogt1990}
{Vogt} N.,  {Barrera} L.~H.,   {Navarro} M.,  1990, \mn@doi [\apss]
  {10.1007/BF00642569}, \href
  {http://adsabs.harvard.edu/abs/1990Ap%26SS.173..145V} {173, 145}

\bibitem[\protect\citeauthoryear{{Waters}, {Cote}  \& {Lamers}}{{Waters}
  et~al.}{1987}]{Waters1987}
{Waters} L.~B.~F.~M.,  {Cote} J.,   {Lamers} H.~J.~G.~L.~M.,  1987, \aap, \href
  {http://adsabs.harvard.edu/abs/1987A%26A...185..206W} {185, 206}

\bibitem[\protect\citeauthoryear{{Wenger} et~al.,}{{Wenger}
  et~al.}{2000}]{Wenger2000}
{Wenger} M.,  et~al., 2000, \mn@doi [\aaps] {10.1051/aas:2000332}, \href
  {http://adsabs.harvard.edu/abs/2000A%26AS..143....9W} {143, 9}

\bibitem[\protect\citeauthoryear{{Zorec} \& {Briot}}{{Zorec} \&
  {Briot}}{1997}]{Zorec1997}
{Zorec} J.,  {Briot} D.,  1997, \aap, \href
  {http://adsabs.harvard.edu/abs/1997A%26A...318..443Z} {318, 443}

\makeatother
\end{thebibliography}

%\begin{thebibliography}{99}
%\end{thebibliography}

%%%%%%%%%%%%%%%%%%%%%%%%%%%%%%%%%%%%%%%%%%%%%%%%%%

%%%%%%%%%%%%%%%%% APPENDICES %%%%%%%%%%%%%%%%%%%%%
\appendix

\newpage
\onecolumn

\section{BeSOS's handbook (online material)}
\label{Ap:besoshandbook}

\section{Table of the derived stellar parameters}
\label{appendix:parte2}
Stellar parameters derived by fitting rotationally convolved stellar model atmospheres (Tlusty) on the lines: HeI $\lambda$4471 and MgII $\lambda$4481, and stellar model atmospheres (Kurucz) on photometric data, are presented here. In total we analyzed 69 Be stars, where 16 are considered as Low Temperatures ($T_\text{eff} < $ 15000K) stars and the results for these stars is based only on the best fit on photometric data (for that reason no $v \, \sin i$ value is given). We note that the deviation percentage can be considered as 1.0$\%$ for the original value for the $T_\text{eff}$ and $\log$ g, and 2.0$\%$ for $R_{\star}$ and $v\sin i$ (see Subsection~\ref{errors}). 

\begin{table}
\scriptsize
\label{tab:results}
\caption{Stellar parameters derived by photometric and spectroscopic analysis}
 \begin{threeparttable}
\begin{tabular}[c]{l  l l | l l l l l l l l}
\multicolumn{1}{c}{ } & \multicolumn{2}{|c|}{Literature} & \multicolumn{8}{|c|}{BeSOS} \\
\hline
Target & Sp.T & D & Sp.T & $T_\text{eff}$ & $\log$ g & $R_{\star}$ & E(B-V) & Av & $v \, \sin i$ & $\chi^{2}$\\
       &    & (pc) &    &  (K)      & (cm s$^{-2}$) & ($R_{\odot}$) &   &  & (km s$^{-1}$)\\ 
\hline             
%HD10144  &  B6Vpe                      & 40000  & 4.0 &   7.40 & 17.50 & 0.12 & 3.1 & 260 \\    
HD 33328  & B2IVne & 248 $\pm$ 11 & B2V & 19526 $\pm$ 195 & 3.30 $\pm$ 0.03 & 7.31 $\pm$ 0.15   & 0.04  & 0.124  & 287 $\pm$ 6 & 4.74 \\  
HD 35165  & B2Vnpe & 440 $\pm$ 106 & B4V & 17000 $\pm$ 170 & 4.00 $\pm$ 0.04 & 5.90 $\pm$ 0.12   & -0.04 & -0.124 & 240 $\pm$ 5 & 3.64\\  
HD 35411  & B1V+B2 & 299 $\pm$ 95 & B0V & 25900 $\pm$ 259 & 4.50 $\pm$ 0.04 & 7.70 $\pm$ 0.15  & 0.02  & 0.062  & 53 $\pm$ 1 & 36.80\\
HD 35439  & B1Vpe & 318 $\pm$ 84 & B1V & 25300 $\pm$ 253  & 3.50 $\pm$ 0.04 & 4.79 $\pm$ 0.10  & 0.04  & 0.124  & 260 $\pm$ 5 & 14.28 \\ 
HD 37041  & O9.5IV & 473 $\pm$ 92 & B0V & 27500 $\pm$ 275 & 3.72 $\pm$ 0.04 & 6.51 $\pm$ 0.13  & 0.02  & 0.062  & 140 $\pm$ 3 & 40.40 \\    
HD 37795\tnote{$^{LT}$} & B9IVe & 80 $\pm$ 2 & B7III & 12200 $\pm$ 122 & 3.50 $\pm$ 0.04 & 7.00 $\pm$ 0.14  & 0.010 & 0.031 & -- & 5.10\\
HD 41335  & B3/5Vne & 403 $\pm$ 121 & B1V & 22500 $\pm$ 225 & 3.35 $\pm$ 0.03 & 7.06 $\pm$ 0.14 & 0.14 & 0.434 & 330 $\pm$ 7 & 14.17 \\ 
HD 42167\tnote{$^{LT}$} & B9IV & 221 $\pm$ 8 & B7III & 11430 $\pm$ 114 & 3.00 $\pm$ 0.03 & 6.75 $\pm$ 0.14  & 0.01 & 0.031 & -- & 3.40 \\  
HD 45725 & B4Veshell & 207 $\pm$ 48 & B2V & 19600 $\pm$ 196 & 3.40 $\pm$ 0.03 & 7.00 $\pm$ 0.14  & 0.04 & 0.124 & 280 $\pm$ 6 & $>$1000 \\ 
HD 45910 & B2IIIe & 752 $\pm$ 158  & B1V & 21300 $\pm$ 213 & 3.94 $\pm$ 0.04 & 9.33 $\pm$ 0.19   & 0.409 & 1.268 & 100 $\pm$ 2 & 69.17 \\  
HD 48917 & B2Ve & 1111 $\pm$ 567  & B1V & 22130 $\pm$ 221 & 3.30 $\pm$ 0.03 & 8.00 $\pm$ 0.16  & 0.06 & 0.186 & 200 $\pm$ 4 & 14.80 \\  
HD 50013 & B1.5Ve & 202 $\pm$ 4 & B0V & 26000 $\pm$ 260 & 3.90 $\pm$ 0.04 & 4.50 $\pm$ 0.09   & 0.02 & 0.062 & 290 $\pm$ 6& 22.10 \\ 
HD 52918 & B1V & 373 $\pm$ 30 & B1V & 25450 $\pm$ 255 & 4.50 $\pm$ 0.04 & 6.50 $\pm$ 0.13   & 0.05 & 0.155 & 242 $\pm$ 5 & 3.68 \\
HD 56014 & B3IIIe & 531 $\pm$ 90 & B2III & 19300 $\pm$ 193 & 3.80 $\pm$ 0.04 & 10.80 $\pm$ 0.22  & 0.00  & 0.00 & 200 $\pm$ 4 & 12.5 \\  
HD 56139 & B2IV-Ve & 279 $\pm$ 13 & B3III & 17170 $\pm$ 172 & 3.89 $\pm$ 0.04 & 10.50 $\pm$ 0.21  &  0.06  & 0.186 & 100 $\pm$ 2 & 8.80 \\
HD 57150 & B2V+B3IVne & 246 $\pm$ 10 & B1V & 22000 $\pm$ 220 & 3.50 $\pm$ 0.04 & 6.90 $\pm$ 0.14   &  0.14 & 0.434 & 180 $\pm$ 4 & 13.40 \\ 
HD 57219 & B3Vne & 232 $\pm$ 12 & B2V & 19300 $\pm$ 193 & 3.99 $\pm$ 0.04 & 4.69 $\pm$ 0.09   & 0.08 & 0.248 & 50 $\pm$ 1 & 4.60 \\
HD 57682 & O9.2IV & 1234 $\pm$ 0 & B0V & 30000 $\pm$ 300 & 4.00 $\pm$ 0.04 & 10.30 $\pm$ 0.21   & 0.11 & 0.341 & 10 $\pm$ 0.2 & 9.30 \\  
HD 58343 & B2Vne & 287 $\pm$ 47 & B2V & 20000 $\pm$ 200 & 3.56 $\pm$ 0.04 & 5.39 $\pm$ 0.11  &  0.17 & 0.527 & 10 $\pm$ 0.2 & 8.37 \\ 
HD 58715\tnote{$^{LT}$} & B8Ve & 49 & B7V & 11820 $\pm$ 118 & 3.20 $\pm$ 0.03 & 3.78 $\pm$ 0.08   & 0.02 & 0.062 & -- & 5.00 \\  
HD 60606 & B2Vne & 363 $\pm$ 31 & B3V & 18000 $\pm$ 180 & 3.53 $\pm$ 0.04 & 6.23 $\pm$ 0.12 & 0.02 & 0.062 & 250 $\pm$ 5 & 46.00 \\ 
%HD 62632\tnote{$^{LT}$} &  Ap & 348 $\pm$ 134 & F6V  & 6400 $\pm$ 64 & 4.20 $\pm$ 0.04 & 2.85 $\pm$ 0.06 & 348  & -0.007 & -0.022 & -- & 9.27 \\
HD 63462 & B1IVe & 434 $\pm$ 43 & B0III & 26000 $\pm$ 260 & 2.90 $\pm$ 0.03 & 13.50 $\pm$ 0.27   & 0.21 & 0.651 & 300 $\pm$ 6 & 11.50 \\  
HD 68423\tnote{$^{LT}$} & B6Ve & 250 $\pm$ 14 & B7V & 12100 $\pm$ 121 & 4.00 $\pm$ 0.04 & 4.11 $\pm$ 0.08  & 0.02 & 0.062 & -- & 3.04 \\  
HD 68980 & B2ne & 284 $\pm$ 12 & B1V & 23668 $\pm$ 237 & 3.32 $\pm$ 0.03 & 6.66 $\pm$ 0.13  & 0.14 & 0.434 & 110 $\pm$ 2 & 10.14 \\
HD 71510 & B2Ve & 213 $\pm$ 10 & B3V & 17643 $\pm$ 176 & 4.10 $\pm$ 0.04 & 4.70 $\pm$ 0.09  & 0.08 & 0.248 & 150 $\pm$ 3 & 4.80 \\
HD 75311 & B3Vne & 185 $\pm$ 4 & B1V & 23895 $\pm$ 239 & 4.00 $\pm$ 0.04 & 3.00 $\pm$ 0.06  & 0.02 & 0.062 & 250 $\pm$ 5 & 13.60 \\ 
HD 78764 & B2IVne & 296 $\pm$ 33 & B3III & 19000 $\pm$ 190 & 3.80 $\pm$ 0.04 & 8.70 $\pm$ 0.17  & 0.09 & 0.279 & 140 $\pm$ 3 & 5.70 \\  
HD 83953 & B5V & 155 $\pm$ 4 & B5V & 15000 $\pm$ 150 & 2.54 $\pm$ 0.03 & 4.00 $\pm$ 0.08   & 0.02 & 0.062 & 250 $\pm$ 5 & 10.80 \\  
HD 89080\tnote{$^{LT}$} & B8IIIe & 104 & B7III & 11600 $\pm$ 116 & 3.50 $\pm$ 0.04 & 7.20 $\pm$ 0.14  & 0.033 & 0.102 & -- & 14.00 \\
HD 89890 & B3IIIe & 352 $\pm$ 34 & B5IV & 15000 $\pm$ 150 & 3.00 $\pm$ 0.03 & 10.07 $\pm$ 0.20  & 0.02 & 0.062 & 26.8 $\pm$ 0.5 & 21.66 \\
HD 91465 & B4Vne & 148 $\pm$ 8 & B3III  & 17410 $\pm$ 174 & 2.96 $\pm$ 0.03 & 7.95 $\pm$ 0.16  & 0.09 & 0.279 & 280 $\pm$ 6 & 14.60 \\
HD 92938 & B3/5V & 139 $\pm$ 3 & B3V & 19000 $\pm$ 190 & 3.48 $\pm$ 0.03 & 3.30 $\pm$ 0.07   & 0.02 & 0.062 & 110 $\pm$ 2 & 14.15 \\
HD 93563 & B8/9III & 161 $\pm$ 4 & B5III & 15000 $\pm$ 150 & 2.30 $\pm$ 0.02 & 6.00 $\pm$ 0.12  & 0.23 & 0.713 & 280 $\pm$ 6 & 15.60 \\  
HD 98058\tnote{$^{LT}$} & A5V & 56 & A4V & 8200 $\pm$ 82 & 3.00 $\pm$ 0.03 & 3.20 $\pm$ 0.06  & 0.02 & 0.124 & -- & 12.60 \\  
HD 102776 & B3Vne & 182 $\pm$ 19 & B2V & 20000 $\pm$ 200 & 3.20 $\pm$ 0.03 & 5.00 $\pm$ 0.10   & 0.02 & 0.062 & 200 $\pm$ 4 & 25.70 \\  
HD 103192 & ApSi & 94 $\pm$ 5 & B8V & 10980 $\pm$ 110 & 2.52 $\pm$ 0.03 & 3.89 $\pm$ 0.08  & 0.00 & 0.00 & -- & 3.94 \\  
HD 105382 & B3/5IIIe & 134 $\pm$ 11 & B5V & 15000 $\pm$ 150 & 3.40 $\pm$ 0.03 & 4.18 $\pm$ 0.08   & 0.02 & 0.062 & 67.5 $\pm$ 1.4 & 12.80 \\ 
HD 105435 & B2Vne & 127 $\pm$ 7 & B1V & 22150 $\pm$ 222 & 3.43 $\pm$ 0.03 & 8.18 $\pm$ 0.16   & 0.15 & 0.465 & 250 $\pm$ 5 & 15.50 \\
HD 107348\tnote{$^{LT}$} & B8V & 127 $\pm$ 3 & B6V & 12830 $\pm$ 128 & 2.26 $\pm$ 0.02 & 4.57 $\pm$ 0.09  & 0.04 & 0.124 & -- & 2.70 \\  
HD 110335\tnote{$^{LT}$} & B5IIIe & 293 $\pm$ 21 & B8IV & 10600 $\pm$ 106 & 2.90 $\pm$ 0.03 & 11.20 $\pm$ 0.22  & 0.02 & 0.062 & -- & 30.00 \\  
HD 110432 & B0.5IVpe & 373 $\pm$ 41 & B1V & 22000 $\pm$ 220 & 3.00 $\pm$ 0.03 & 6.00 $\pm$ 0.12   & 0.02 & 0.062 & 400 $\pm$ 8 & 70.9 \\  
HD 112078 & B3Vne & 117 $\pm$ 2 & B4V & 16500 $\pm$ 165 & 3.01 $\pm$ 0.03 & 3.00 $\pm$ 0.06   & 0.03 & 0.093 & 290 $\pm$ 6 & 4.46\\
HD 120324 & B2Vnpe & 155 $\pm$ 3 & B1V & 23950 $\pm$ 240 & 4.00 $\pm$ 0.04 & 5.37 $\pm$ 0.11  & 0.02 & 0.062 & 110 $\pm$ 2 & 20.20 \\ 
HD 121492\tnote{$^{LT}$} & K0 & 505 $\pm$ 262 & K2III & 4736 $\pm$ 47 & 2.00 $\pm$ 0.02 & 18.65 $\pm$ 0.37  & 0.40 & 1.244 & -- & 20.75 \\ 
HD 124195 & B5V & 315 $\pm$ 35 & B5V & 15000 $\pm$ 150 & 4.00 $\pm$ 0.04 & 3.80 $\pm$ 0.08 & 0.02 & 0.062 & 150 $\pm$ 3 & 36.80 \\
HD 124367 & B4Vne & 147 $\pm$ 5 & B2V & 19650 $\pm$ 197 & 3.00 $\pm$ 0.03 & 3.10 $\pm$ 0.06  & 0.02 & 0.062 & 260 $\pm$ 5& 51.60 \\  
HD 124771 & B4V & 197 $\pm$ 8 & B4V & 17100 $\pm$ 171 & 3.50 $\pm$ 0.04 & 4.30 $\pm$ 0.09   & 0.07 & 0.217 & 150 $\pm$ 3 & 9.60 \\
HD 127972 & B2Ve & 93 $\pm$ 1 & B1V & 21000 $\pm$ 210 & 3.95 $\pm$ 0.04 & 6.10 $\pm$ 0.12 & 0.02 & 0.062 & 240 & 25.40 \\  
HD 131492 & B2IV/Ve & 462 $\pm$ 62 & B2III & 20000 $\pm$ 200 & 4.00 $\pm$ 0.04 & 10.80 $\pm$ 0.22 & 0.18 & 0.558 & 100 $\pm$ 2 & 11.70 \\  
HD 135734\tnote{$^{LT}$} & B8Ve & 102 $\pm$ 7 & B6V & 13500 $\pm$ 135 & 3.80 $\pm$ 0.04 & 3.80 $\pm$ 0.08   & 0.03 & 0.093 & -- & 26.70 \\  
HD 138769 & B3IVp & 131 $\pm$ 7 & B3V & 18400 $\pm$ 184 & 3.50 $\pm$ 0.04 & 3.00 $\pm$ 0.06  & 0.02 & 0.062 & 30 $\pm$ 0.6 & 58.50 \\
HD 142184 & B2V & 130 $\pm$ 7 & B3V & 17375 $\pm$ 174 & 4.25 $\pm$ 0.04 & 3.03 $\pm$ 0.06   & 0.18 & 0.558 & 300 $\pm$ 6 & 4.48 \\
HD 142983 & B8Ia/Iab & 143 $\pm$ 4 & B3V & 18000 $\pm$ 180 & 3.71 $\pm$ 0.04 & 3.11 $\pm$ 0.06   & 0.08 & 0.257 & 370 $\pm$ 7 & 12.90 \\  
HD 143275 & B0.3IV & 150 $\pm$ 20 & O9V & 30900 $\pm$ 309 & 3.50 $\pm$ 0.04 & 7.34 $\pm$ 0.15   & 0.17 & 0.527 & 257 $\pm$ 5 & 13.80 \\  
HD 148184 & B2Vne & 161 $\pm$ 5 & B0V & 30000 $\pm$ 300 & 3.53 $\pm$ 0.04 & 4.44 $\pm$ 0.09   & 0.40 & 1.240 & 150 $\pm$ 3 & 64.00 \\  
HD 157042 & B2Vnne & 286 $\pm$ 21 & B1V & 22000 $\pm$ 220 & 3.90 $\pm$ 0.04 & 5.17 $\pm$ 0.10   & 0.12 & 0.372 & 280 $\pm$ 6 & 11.80 \\  
HD 157246 & B1Ib & 341 $\pm$ 18 & B1III & 21240 $\pm$ 212 & 3.50 $\pm$ 0.04 & 14.20 $\pm$ 0.28   & 0.08 & 0.248 & 230 $\pm$ 5 & 4.70 \\  
HD 158427 & B2Vne & 81 $\pm$ 5 & B3V & 18110 $\pm$ 181 & 3.80 $\pm$ 0.04 & 5.07 $\pm$ 0.10  & 0.07 & 0.217 & 285 $\pm$ 6 & 15.70 \\ 
HD 167128 & B3II/III & 233 $\pm$ 14 & B2V & 20000 $\pm$ 200 & 3.80 $\pm$ 0.04 & 3.90 $\pm$ 0.08   & 0.14 & 0.434 & 50 $\pm$ 1 & 10.00\\  
HD 181615\tnote{$^{LT}$} & F2p & 546 $\pm$ 69  &  &   &  &  &  &   &   &  \\ 
HD 205637 & B3V & 323 $\pm$ 18 & B2V & 20000 $\pm$ 200 & 4.10 $\pm$ 0.04 & 7.00 $\pm$ 0.14   & 0.02 & 0.062 & 230 $\pm$ 5 & 24.00 \\  
HD 209014\tnote{$^{LT}$}  &  B8/9V+B8/9 & 250 $\pm$ 30 & B7III & 12200 $\pm$ 122 & 3.30 $\pm$ 0.03 & 6.00 $\pm$ 0.12  & 0.03 & 0.093 & -- & 6.10 \\ 
HD 209409 & B7IVe & 133 $\pm$ 4 & B5V & 15000 $\pm$ 150 & 3.00 $\pm$ 0.03 & 2.90 $\pm$ 0.06   & 0.03 & 0.093 & 350 $\pm$ 7 & 15.60 \\ 
HD 212076 & B2IV-Ve & 497 $\pm$ 69 & B3III & 18000 $\pm$ 180 & 4.00 $\pm$ 0.04 & 11.60 $\pm$ 0.23  & 0.02 & 0.062 & 100 $\pm$ 2 & 20.35 \\
\hline
\end{tabular}
\begin{tablenotes}
      \item[$^{LT}$] means Low Temperature stars, where the best fit was found only for photometric data.
      \item[] $\chi^{2}$ is the reduced Chi-square value obtained for the best fit on photometric data.
      \item[] Information from literature, Sp.T and D, are taken from Simbad database.
\end{tablenotes}
\end{threeparttable}
\end{table}

\begin{table}
\scriptsize
 \contcaption{} 
 \begin{threeparttable}
\begin{tabular}[c]{l  l l | l l l l l l l l }
\multicolumn{1}{c}{ } & \multicolumn{2}{|c|}{Literature} & \multicolumn{8}{|c|}{BeSOS} \\
\hline
Target & Sp.T & D & Sp.T & $T_\text{eff}$ & $\log$  g & $R_{\star}$ & E(B-V) & Av & $v \, \sin i$ & $\chi^{2}$\\
       &    & (pc) &    &  (K)      & (cm s$^{-2}$) & ($R_{\odot}$) &   &  & (km s$^{-1}$)\\ 
\hline 
HD 212571 & B1III-IVe & 239 $\pm$ 16 & B1V & 24500 $\pm$ 245 & 3.40 $\pm$ 0.03 & 5.90 $\pm$ 0.12  & 0.22 & 0.682 & 215 $\pm$ 4 & 15.62 \\
HD 214748\tnote{$^{LT}$} & B8Ve & 149 $\pm$ 14 & B7III & 11500 $\pm$ 115 & 3.90 $\pm$ 0.04 & 6.70 $\pm$ 0.13   & 0.01 & 0.031 & -- & 3.90 \\
HD 217891 & B6Ve & 125 $\pm$ 3 & B4V & 16000 $\pm$ 160 & 3.00 $\pm$ 0.03 & 3.70 $\pm$ 0.07   & 0.04 & 0.124 & 90 $\pm$ 2 & 22.60 \\  
HD 219688\tnote{$^{LT}$} & B7/B8V & 123 $\pm$ 5 & B5V & 14200 $\pm$ 142 & 3.86 $\pm$ 0.04 & 3.88 $\pm$ 0.08   & 0.00 & 0.00 & -- & 3.70 \\  
HD 221507\tnote{$^{LT}$} & B9.5IIIpHgMgSi & 53 & B8V & 11000 $\pm$ 110 & 4.40 $\pm$ 0.04 & 2.00 $\pm$ 0.04  & -0.03 & 0.093 & -- & 2.80 \\  
HD 224686\tnote{$^{LT}$} & B8V & 114 $\pm$ 2 & B9V & 10500 $\pm$ 105 & 3.35 $\pm$ 0.03 & 4.30 $\pm$ 0.09  & 0.02 & 0.062 & -- & 4.58 \\
  \hline
 \end{tabular}
\begin{tablenotes}
      \item[$^{LT}$] means Low Temperature stars, where the best fit was found only for photometric data.
      \item[] $\chi^{2}$ is the reduced Chi-square value obtained for the best fit on photometric data.
      \item[] Information from literature, Sp.T and D, are taken from Simbad database.
\end{tablenotes}
\end{threeparttable}
\end{table}

%%%%%%%%%%%%%%%%%%%%%%%%%%%%%%%%%%%%%%%%%%%%%%%%%%

% Don't change these lines
\bsp	% typesetting comment
\label{lastpage}

\end{document}